\documentclass{article}[12pt]
\usepackage{graphicx}
\usepackage{setspace}
\usepackage{amsmath}

 \DeclareGraphicsExtensions{.eps, .ps}
\font\fiverm=cmr5

\def\pr{\mathop{\rm pr}\nolimits}
\def\cov{\mathop{\rm cov}\nolimits}
\def\var{\mathop{\rm var}\nolimits}

\def\Tmax{T_{\hbox{\fiverm max}}}
\def\indep{\mathrel{\rlap{$\perp$}\kern1.6pt\mathord{\perp}}}
\newbox\dashbox
\setbox\dashbox=\vbox to 6pt{\vfill\hbox to 0.4cm{\hrulefill}\vfill}

\def\rip{\flat}
\def\H{{\cal H}}

\def\Rsharp{R^\sharp}
\def\U{{\cal U}}
\def\V{{\cal V}}

\def\given{\mathrel{|}}
\def\bft{{\bf t}}
\def\bfs{{\bf s}}

\def\Record{{\cal R}}

\def\one{{\mathbf 1}}
\def\vhalf{\textstyle{\frac12}}
\mathchardef\Real="023C

\def\normsq#1{\Vert#1\Vert^2}
\newif\ifignoretext \ignoretexttrue
\def\beginignoretext{\setbox0=\vbox\bgroup}
\def\endignoretext{\egroup \ifignoretext\relax\else\unvbox0 \fi}
\def\acknowledgement{\bigbreak\leftline{\large \bf Acknowledgements}\nobreak\smallskip\noindent}
\newbox\bigstrutbox
\setbox\bigstrutbox=\hbox{\vrule height 10.5pt depth3.5pt width 0pt}
\def\bigstrut{\relax\ifmmode\copy\bigstrutbox\else\unhcopy\bigstrutbox\fi}
\chardef\tie='176
\chardef\caret='136

\begin{document}
%\title{Revival models and survival studies}
\title{Survival models and health sequences}
\author{Walter Dempsey and Peter McCullagh\thanks{Department of Statistics,
University of Chicago, 5734
University Ave, Chicago, Il 60637, U.S.A.}}
\maketitle
\begin{abstract}\noindent
Survival studies often generate not only a survival time for each patient but also a sequence of
health measurements at annual or semi-annual check-ups while the patient remains alive.
Such a sequence of random length accompanied by a survival time is called a survival process.
Ordinarily robust health is associated with longer survival, so the two parts
of a survival process cannot be assumed independent.
This paper is concerned with a general technique---reverse alignment---for
constructing statistical models for survival processes.
A revival model is a regression model in the sense that it incorporates
covariate and treatment effects into both the
distribution of survival times and the joint distribution of health outcomes.
The revival model also determines a conditional survival distribution given the observed history,
which describes how the subsequent survival distribution
is determined by the observed progression of health outcomes.
\end{abstract}


\section{Survival studies}
A survival study is one in which
patients are recruited according to well-defined selection criteria
and their health status monitored on a regular or intermittent schedule
until the terminal event, here assumed to be fatal.
Covariates such as sex and age are recorded at the time of recruitment,
and, if there is more than one treatment level, the assignment is
presumed to be randomized.
In a simple survival study, the health status $Y(t)$ at time~$t$ is a
binary variable, dead or alive, and the entire process is then
summarized by the time $T>0$ spent in state~1, i.e.,~the survival time.
In a survival study with health monitoring, $Y(t)$~is a more detailed
description of the state of health or quality of life of the individual,
containing whatever information---pulse rate, cholesterol level,
cognitive score or CD4 cell count---is deemed relevant to the study.
The goal may be to study the effect of treatment on survival time,
or to study its effect on quality of life,
or to predict the subsequent survival time of patients 
given their current health history.

Survival studies with intermittent health monitoring are moderately common,
and likely to become more so as health records become available electronically
for research purposes.
Within the past few years, several issues of the journal {\it Lifetime Data Analysis} have
been devoted to problems connected with studies of exactly this type.
For a good introduction, with examples and a discussion of scientific objectives,
see Diggle, Sousa and Chetwynd (2006),
Kurland, Johnson, Egleston and Diehr (2009)
or Farewell and Henderson (2010).
Section~8 of van Houwelingen and Putter (2012)
is recommended reading.

In practice, the patient's health status is measured at recruitment ($t=0$),
and regularly or intermittently thereafter while the patient remains alive.
To emphasize the distinction between the observation times and observation values,
each time is called an appointment date, the set of dates is called the
appointment schedule.
Apart from covariate and treatment values, a complete uncensored observation on one
patient $(T, \bft, Y[\bft])$ consists of the survival time $T > 0$,
the appointment schedule $\bft \subset [0, T)$,
and the health status measurements $Y[\bft]$ at these times.
To accommodate patients whose record is incomplete, a censoring indicator
variable is also included.
In that case, the censoring time is usually, but not necessarily,
equal to the date of the most recent appointment.

%A statistical model for a survival study is a family of probability
%distributions for the record of each patient, all three components included.
%At a minimum, therefore, it is necessary to model the survival time
%and the state of health jointly,
%and to consider how the joint distribution might be affected by treatment.
%Ordinarily, robust health is associated with longevity, but
%if both are affected by treatment, there is no guarantee that
%the two effects are in similar  directions.

In the sense that the health status is measured over time on each patient,
a survival study is a particular sort of longitudinal study.
Certainly, temporal and other correlations are expected and must be accommodated.
But the distinguishing feature, that each sequence is terminated by failure or censoring,
gives survival-process models a very distinct character:
as an absorbing state, death, contradicts stationarity.
For a good survey of the goals of such studies and the modeling strategies employed, see
Kurland, Johnson, Egleston and Diehr (2009).

The goal of this paper is not so much to recommend a particular statistical model,
as to explore a general mathematical framework 
for the construction of survival-process models,
permitting easy computation of the likelihood function and parameter estimates,
and straightforward derivation of predictive distributions for individual survival times.
For example, the paper has nothing to say on the choice between proportional hazards and
accelerated lifetimes for accommodating treatment effects.
Apart from reservations concerning the use of time-evolving covariates,
all standard survival models are acceptable within the framework.
Nor has the paper anything to contribute to the choice between
Bayesian and non-Bayesian methods of analysis;
prior distributions are not discussed, so either approach can be used.
Administrative complications of the sort that are inevitable in medical
and epidemiological research will be
ignored for the most part, so no attempt is made to provide a complete turnkey package.
For example, the paper has little to say about how best to handle incomplete records
other than to recognize that censoring and delayed reporting are issues that must be addressed---again
using standard well-developed methods.
Since most of the computations needed for model fitting and parameter estimation
are relatively standard and need not involve specialized Markov chain or Monte Carlo algorithms,
detailed discussion of computational techniques is omitted.
The emphasis is on statistical ideas and principles, 
strategies for model formulation, sampling schemes, and
the distinction between time-dependent variables and time-evolving variables
in the definition of treatment effects.

\section{Reverse alignment}
\subsection{The survival process}
A survival process $Y$ is a stochastic process defined for real~$t$,
in which $Y_i(t)$ is the state of health or quality of life of
patient~$i$ at time~$t$, usually measured from recruitment.
%In this and in subsequent sections $Y$ denotes an observable process.
In a simple survival process, the state space $\Record=\{0, 1\}$
is sufficient to encode only the most basic of vital signs, dead or alive;
more generally, the state space is any set large enough to encode
the observable state of health or quality of life of the patient at one instant in time.
Flatlining is the distinguishing characteristic of a survival process,
i.e.,~$\rip\in\Record$ is an absorbing state such that
$Y(t) = \rip$ implies $Y(t') = \rip$ for all $t'\ge t$.
The survival time is the time to failure:
$$
T_i = \sup_{t \ge 0} \{t : Y_i(t) \neq \rip\};
$$
it is presumed that $Y_i(0)\neq\rip$ at recruitment, so $T_i > 0$.
This definition is quite general, and does not exclude immortality,
i.e.,~$T = \infty$ with positive probability.
In all of the models considered here, however,
survival  time is finite with probability one.

\subsection{Administrative and other schedules}\label{schedules}
Since the appointment schedule is a random subset $\bft \subset [0, T)$,
it is obviously informative for survival: $T > \max(\bft)$.
If better health is associated with longer survival,
we should expect patients who are initially frail to have shorter
health records than patients who are initially healthy.
In other words, even if the trajectories for distinct
individuals may be identically distributed,
the first component of a short health-status record
should not be expected to have the same distribution as
the first component of a longer record.
On the contrary, any model such that record length is independent of
record values must be regarded as highly dubious for survival studies.
It is necessary, therefore, to address the nature of the information contained in~$\bft$.

Consider a patient who has had appointments on $k$ occasions
$\bft^{(k)} = (t_0<\cdots<t_{k-1})$.
The sequence $Y[\bft^{(k)}]$ of recorded health values 
may affect the scheduled date $t_k$ for the next appointment:
for example, patients in poor health needing more careful monitoring
may have short inter-appointment intervals.
Whatever the scheduled date may be, the appointment is null
unless $t_k < T$.
The assumption used in this paper is sequential conditional independence,
namely that
\begin{equation}\label{sci}
t_k \indep Y \given (T, \bft^{(k)}, Y[\bft^{(k)}]).
\end{equation}
In other words, the conditional distribution of the random interval $t_k - t_{k-1}$  
may depend on the observed history $Y[\bft^{(k)}]$, 
but not on the subsequent health trajectory except through~$T$.
Here, $t_k$~may be infinite (or null) with positive probability,
in which case the recorded sequence is terminated at $t_{k-1}$.

The schedule is said to be \emph{administrative} if $t_k$ is a deterministic function
of the pair $(\bft^{(k)}, Y[\bft^{(k)}])$, implying that the conditional distribution (\ref{sci}) is degenerate.
Eventually, for some finite~$k$, the patient dies or is censored at time $T\in (t_{k-1} , t_k)$ 
while the next appointment is pending, 
so the recorded schedule is $\bft = \bft^{(k)} = \bft^{(k+1)} \cap[0, T)$.
Equivalently, the last recorded value is $(t_k, \rip)$.

%\beginignoretext
%In constructing a probability distribution for the record of one patient,
%it is essential to bear in mind that the three components
%$(T, \bft, Y[\bft])$ cannot be independent.
%First, $T>0$ is a positive random variable, and the appointment schedule
%$\bft$ is a finite subset of the random interval $[0, T)$,
%which implies that $\bft$ is also a random variable.
%Moreover $T > \max(\bft)$ implies that the appointment schedule
%for any patient is informative about his or her survival time.
%Likewise, since the number of health-status measurements coincides
%with $\#\bft$, it also should be strongly correlated with survival.
%Second, 
% 
%Despite these complications, we aim to construct a statistical model
%that has the right sorts of symmetries and is not self-contradictory
%or otherwise inappropriate in any of the senses discussed above.
%Progress in this direction requires a few assumptions, and
%in this paper, the first assumption is that
%\begin{equation}%\label{condindep}
% \bft \indep Y \given T.
%\end{equation}
%This condition does not require appointments to be made regularly or
%kept sedulously, but it does demand that the probability of an appointment being
%missed while the patient lives should not depend on the state of health,
%except through~$T$.
%A similar assumption is made explicitly or implicitly by most authors:
%see Henderson, Diggle and Dobson (2000, section~2.1).
%The assumption is mathematically natural, and is automatically satisfied if
%$Y$~is binary,  but it is not one to be taken for granted otherwise.
%\endignoretext

While the sequential conditional independence assumption is mathematically clear-cut,
the situation in practice may be considerably more muddy.
Consider, for example, the CSL1 trial
organized by the Copenhagen Study Group for Liver Diseases in the 1960s to study
the effect of prednisone on the survival of patients diagnosed with liver cirrhosis.
In this instance $Y(\cdot)$ is a composite blood coagulation index called the
prothrombin level:
details can be found in Andersen, Hansen and Keiding (1991).
Beginning at death, the reverse-time mean intervals between appointments are
$77$, $210$ and $252$ days, while the medians are $21$, $166$ and $293$ days.
%0.211, 0.244, 0.304 years, while the medians are 0.058, 0.222 and 0.275 years.
In other words, half of the patients who died had their final appointment within
the last three weeks of life.
It is evident that the appointment intensity increases as $s\to 0$ in reverse time,
which is not, in itself, a violation of~(\ref{sci}).
However, one might surmise that the increased intensity is related to
the patient's state of health or perception thereof.
Condition~(\ref{sci}) implies that the appointment intensity does not depend on the 
blood coagulation index other than at earlier appointments,
and it is then unclear to what extent the condition may be violated by
patient-initiated appointments.
Liest\o l and Andersen (2002, section~4.1) note that 71 off-schedule appointments
occurred less than 10 days prior to death, the majority of which were patient-initiated.
They also examine the effect on hazard estimates of
excluding unscheduled prothrombin measurements.
%
% r1 <- tapply(revival, id, min)
% min2 <- function(x){diff(sort(x))[1]};  r2 <- tapply(revival, id, min2)
% min3 <- function(x){diff(sort(x))[2]};  r3 <- tapply(revival, id, min3)
% min4 <- function(x){diff(sort(x))[3]};  r4 <- tapply(revival, id, min4)
% summary(r1*365);  summary(r2*365); summary(r3*365);  summary(r4*365)
%   Min. 1st Qu.  Median    Mean 3rd Qu.    Max.     NAs 
%   0.00    3.00   21.00   76.86   88.25  762.00 
%  15.00   90.25  165.50  209.80  350.00 1097.00      22 
%   35.0   106.0   292.5   251.6   365.5   726.0      56 
%   28.0   104.0   238.0   241.8   357.5   763.0      92 
%t1 <- tapply(time, id, min);  t2 <- tapply(time, id, min2);  t3 <- tapply(time, id, min3)
%summary(t1*365);  summary(t2*365); summary(t3*365);  summary(t4*365)
%   Min. 1st Qu.  Median    Mean 3rd Qu.    Max.    NA's 
%      0       0       0       0       0       0 
%     15      88      96     103     110     446      22 
%   15.0    88.0    94.0   107.8   105.3   525.0      56 
%   28.0   104.0   238.0   241.8   357.5   763.0      92 
%

Although we refer to $Y(\cdot)$ generically as the patient's {\it state of health,}
this description is not to be taken literally.
The actual meaning depends on what has in fact been measured:
in general, $Y(\cdot)$~is only one component or one aspect of patient health.

\subsection{The revival process}
On the assumption that the survival time is finite,  the time-reversed process
$$
Z_i(s) = Y_i(T_i - s)
$$
is called the revival process.
Thus, $Z_i(s)$ is the state of health of patient~$i$ at time $s$ prior to failure,
and $Z_i(T_i) = Y_i(0)$ is the value at recruitment.
By construction, $Z(s) = \rip$ for $s < 0$, and $Z(s)\neq \rip$ for $s > 0$.
Although $Z$ is defined in reverse time, the temporal evolution via
the survival process occurs in real time:
by definition, $Z(\cdot)$~is not observable component-wise until the patient dies.
The transformation $Y \mapsto (T, Z)$ is clearly invertible;
it may appear trivial, and in a sense it is trivial.
Its one key property is that the revival process $Z$ and the random variable $T$
are variation independent. 
%In the statistical models considered here, variation independence may be exploited
%through the revival assumption, which states that the revival process
%and the survival time are statistically independent.
%More generally, $Z \indep T \given X$ if covariates are present.
%The revival assumption provides a convenient starting point for model construction;
%it is not critical to any part of the theory presented here.

%To understand what the revival assumption implies,
%consider two patients $i, j$ with identical covariate values $x_i=x_j$,
%whose survival times are $T_i=5$ and $T_j=20$ time units respectively.
%Exchangeability implies that their health status values at revival time~$s$,
%$Z_i(s)$ and $Z_j(s)$, are identically distributed,
%and the revival assumption implies that both are independent of the survival times.
%The revival assumption is trivially satisfied by survival models
%with simple follow-up, where $Y(t)$ is binary.
%
%The revival assumption provides a convenient starting point for model construction;
%it is not critical to any part of the theory presented here.
%The more fundamental assumption is that the appointment schedule
% should be conditionally independent of the revival process given the survival time,
% i.e.,~$\bft \dash T \dash Z$ in standard graphical-model notation.
% The revival assumption is a simplification implying that
%the edge connecting $Z$ with $T$ is omitted.
% Covariates, if there are any, are regarded as a fixed function on the patients.

\begin{table}[ht] 
\centering
\caption{Average prothrombin levels indexed by $T$ and $t$} 
\begin{tabular}{c|ccccccccc}
Survival&\multicolumn{9}{c}{Time $t$ after recruitment (yrs)\bigstrut} \\ 
time ($T$) &0--1 &1--2 &2--3 &3--4 &4--5 &5--6 &6--7 &7--8   &8+ \cr
\noalign{\hrule\smallskip}
0--1 &58.0 \cr
1--2 &72.5 &66.4 \cr
2--3 &72.6 &73.2 &66.0 \cr
3--4 &69.8 &71.2 &68.5 &54.2 \cr
4--5 &68.5 &75.7 &72.5 &74.6 &57.7 \cr
5--6 &70.5 &77.3 &73.5 &57.1 &64.5 &60.9 \cr
6--7 &81.8 &73.6 &81.1 &80.6 &79.4 &75.5 &75.8 \cr
7--8 &84.4 &88.8 &88.1 &92.1 &85.2 &81.2 &84.3 &88.1 \cr
8+   &77.3 &73.6 &87.0 &74.1 &92.0 &80.3 &89.2 &79.4 &84.7\cr
\hline
\end{tabular}
\label{avgproth:tab}
\end{table}

The chief motivation for time reversal has to do with the effective alignment of patient records for
comparison and signal extraction.
Are the temporal patterns likely to be more similar in two records aligned either by patient age
or by recruitment date,
or are they likely to be more similar in records aligned by reverse age (time remaining to failure)?
Ultimately, the answer must depend on the context, but the 
context of survival studies suggests that the latter may be more effective than the former.
Table~\ref{avgproth:tab} shows the averaged $Y$-values indexed by $T$ and $t$ for the prothrombin example
discussed in more detail in section~\ref{cirrhosis_example}.
It should be borne in mind that each cell is the average of 8--266 non-independent 
high-variability measurements, the larger counts occurring in the upper left cells.
Alignment by reverse time is equivalent to counting leftwards from the main diagonal.
Despite certain anomalies in the table of averages, e.g.~row~6,~column~4,
it is clear that reverse-time is a more effective way of organizing the data
to display the main trends in the mean response:
the forward- and reverse-time sums of squares (equally weighted)
are $543.0$ and $1132.8$ respectively, both on eight degrees of freedom.

Further confirmation is provided in Table~\ref{anova:tab}, which shows
the output from a standard equally-weighted analysis of variance
applied to the table of averages, with three
factors, row, column and diagonal (reverse time), denoted by $R$, $C$ and~$D$ respectively.
Compared with the residual mean square of 23.7, there is
considerable excess variation associated with rows (116.8)
and with the reverse-time factor (77.8), but not so much with columns (34.0).
In other words, the means in Table~1 are expressible
approximately as $\alpha_T+\beta_{T-t}$.
Figures~8.3 and~8.4 of van Houwelingen and Putter (2012),
which are not substantially different from Fig.~\ref{fig:mean_traj} of this paper,
offer strong confirmation of this viewpoint in one further survival study involving
white blood cell counts for patients suffering from chronic myeloid leukemia.
For an application unrelated to survival,
see example~B of Cox and Snell (1981).

\begin{table}[ht]
\centering
\caption{ANOVA decomposition for Table~\ref{avgproth:tab}}
\begin{tabular}{lccrr}
Source &$\U/\V$ & $\normsq{P_\U Y} - \normsq{P_\V Y}$ & d.f. & M.S.\bigstrut\\
\hline
Diagonal&$(R+C+D)/(R+C)$ & 544.3 & 7 & 77.8 \\
Column&$(R+C+D)/(R+D)$ & 237.9 & 7 & 34.0 \\
Row&$(R+C+D)/(C+D)$ & 817.3 & 7 & 116.8 \\
Residual&$RC/(R+C+D)$  & 497.2 & 21 & 23.7\\
\hline
\end{tabular}
\label{anova:tab}
\end{table}

\subsection{Covariates}\label{covariates}
In the absence of specific information to the contrary, responses 
for distinct units are presumed to be identically distributed.
In the great majority of situations, specific information does exist
in the form of covariates or classification variables or relationships.
A covariate is a function $i\mapsto x_i$ on the units, in principle known for all units
whether they occur in the sample or not.
A covariate implies a specific form of inhomogeneity such that equality of covariates
implies equality of response distributions:
$x_i=x_j$ implies $Y_i\sim Y_j$.
In practice, approximate equality of $x$-values also implies
approximate equality of distributions.
Likewise, a relationship is a function on pairs of units such that
$R(i,i') = R(j, j')$ implies $(Y_i, Y_{i'}) \sim (Y_j, Y_{j'})$
for distinct pairs $i\neq i'$, $j\neq j'$, provided that the two pairs
also have the same covariate values:
$(x_i, x_{i'}) = (x_j, x_{j'})$.
Geographic distance and genetic distance are two examples of symmetric relationships.
The overarching principle is that differences in distribution,
marginal or joint, must be associated with
specific inhomogeneities in the experimental material.

The status of certain variables in specific survival studies may appear genuinely unclear.
The conventional rationalization,
in which certain variables used for prediction are notionally `fixed' or non-random
and treated as covariates, is not especially helpful.
Consider, for example, marital status as one variable in a geriatric study
in which the goal is to study both quality of life and survival time.
However it is defined, quality of life is a multi-dimensional response,
a combination of mobility, independence, optimism, happiness, family support,
and so forth.
Marital status is a temporal variable known to be associated with
survival and with quality of life;
one goal may be to predict survival given marital status,
or even to recommend a change of status in an effort to improve the quality of life.
Another example of a similar type is air quality and its relation to
the frequency and severity of asthmatic attacks
(Laird, 1996).
Should such a variable be regarded as a covariate or as one component
of the response?
For survival studies, and for longitudinal studies generally,
the answer is clear:
\emph{every time-evolving variable 
is necessarily part of the response process.}

By definition, a temporal variable $x$ is a function defined for every $t \ge0$.
A temporal variable is a covariate if it is also a function on the units,
meaning that the entire function is determined and recorded at baseline.
Usually this means that $x$~is constant in time, but there are exceptions
such as patient age: see also section~\ref{treatment}.
Marital status and air quality, however, are not only temporal variables,
but variables whose trajectories evolve over real time;
neither is available as a covariate at baseline.

With marital status as a component of the survival process, the joint distribution may
be used to predict the survival
time beyond~$t$ of an individual whose marital history and other health-status
measurements at certain times prior to~$t$ are given.
For that purpose, it is necessary to compute the conditional distribution
of~$T\!$, or more generally of~$Y\!$, given the observed history $\H_t$
at the finite set of appointments prior to~$t$.
For such calculations to make mathematical sense,
marital status must be a random variable, a function of the process~$Y\!$.
Thus, the statement `marital status is a random process' is not 
to be construed as a sociological statement
about the fragility of marriage or the nature of human relations;
it is merely a mathematical assertion to the effect that
probabilistic prediction is not
possible without the requisite mathematical structure of
$\sigma$-fields $\H_t \subset \H_{t'}$ for $t \le t'$
and probability distributions.

\subsection{Treatment}\label{treatment}
A treatment arm is a protocol specifying the therapy, drug type, dose level,
manner of ingestion, and even the next appointment date, as a function
of current medical circumstances and health history.
Examples of simple treatment arms include
one-time surgical procedures with follow-up care as appropriate,
or a fixed pharmaceutical regimen such as 10 mg.~Lipitor per day,
or regular attendance at weekly counselling sessions.
In general, a treatment arm may specify a range of different actions
depending on current health and past history,
so two individuals on the same arm need not be experiencing
the same medical therapy at the same time.

Treatment refers to a scheduled intervention or series of interventions
in which, at certain pre-specified times following recruitment,
patient~$i$ is switched from one arm to another.
Thus, $a_i(t)$ is the treatment arm scheduled for patient~$i$ at time~$t\ge0$.
In general, but crucially  for revival models,
a null level is needed for $t\le0$, including the baseline $t=0$.
The entire temporal trajectory $a_i(t)$ for $t>0$ is determined
by randomization and recorded at baseline.
It does not evolve over real time in response to the
doctor's orders or the patient's perceived needs,
so it is not a time-evolving variable.
Ordinarily, the random variables $a_1(\cdot), \ldots, a_n(\cdot)$ are not independent.
In the sense that it is recorded at baseline, $a_i(\cdot)$ is a covariate;
in the sense that it is a temporal function, it is a time-dependent covariate.

Apart from crossover trials, the distribution of $a(\cdot)$ is such that a switch 
of treatment arms occurs only once,
and then only immediately after recruitment.
Nonetheless, more general formulation is retained to underline
the fact that treatment is a scheduled intervention such that $a_i(t) \neq a_i(0)$,
and thus not constant in time.
Unlike the survival process, the treatment schedule does not evolve randomly in real time.

%It should be understood that a treatment arm is a protocol,
%also called a dynamic treatment regime (Murphy, 2003),
%specifying  the drug type, dose level, frequency, manner of ingestion,
%and even the next appointment date,
%as a function of the current medical circumstances and health history.
%A treatment arm is not necessarily associated with a specific drug
%or a specific medical therapy.
%Consider a hypertension study in which  blood pressure $Y(\cdot)$ in conventional units
%is measured at regular appointments.
%Thus, {\it one blue pill to be taken three times daily while blood pressure exceeds 180,
%and one white pill twice daily if the pressure is between 160 and 180\/}
%is a treatment arm in which the actual therapy and dose level at time $t'$ depend on the
%outcome $Y(t)$ at the most recent appointment $t \le t'$.
%Another treatment arm might reverse the colours or adjust the doses or change the
%therapy or reduce the inter-appointment interval if preceding $Y$-values are not encouraging.
%In this setting, $a_i(t)$ denotes the assigned treatment arm, not the current drug or the dose,
% so $a_i(\cdot)$ is ordinarily constant for $t > 0$,
%even for dynamic chemotherapy strategies such as those discussed by
%Rosth\o j, Keiding and Schmiegelow, (2012).
%For a compliant patient, the drug type and ingested dose can, in principle, be determined from the
%treatment arm and previously recorded $Y$-values.
%The key point is that each treatment arm be fully specified in advance,
%and the assignment be randomized at recruitment.

Let $\bar a_i(s) = a_i(T_i - s)$ be the treatment arm expressed in revival time,
so that, in the standard setting, $\bar a_i(s)$ is null for $s\ge T_i$.  
While $a_i (\cdot)$ is a covariate, $\bar{a}_i (\cdot)$ is not.
It is automatic that that $Z \indep T \given \bar a$,
because $T$ is a function of~$\bar a$.
In the case of treatment, however, the crucial assumption is {\it lack of interference},
i.e.,~the treatment assigned to one individual has no effect on the
response distribution for other individuals,
and the treatment protocol at one point in time has no
effect on the response distribution at other times.
For the latter, the statement is as follows.
For each finite subset $\bfs\subset\Real^+$,  the conditional
distribution of $Z[\bfs]$ given the treatment schedule and survival time
depends only on the treatment arms $\bar a[\bfs]$ prevailing at the scheduled
times,~i.e.,
\[
Z[\bfs] \indep \bar a \given \bar a[\bfs].
\]
For crossover trials in particular, this is a strong assumption denying carry-over effects 
from earlier treatments or later treatments.
It implies in particular that $Z(s) \indep T \given \bar a(s)$,
which is primarily a statement about the one-dimensional marginal distributions.
Note, however, that the interference assumption is relatively benign if $a_i(t)$ is constant
for $t > 0$, as is ordinarily the case.

It is common practice in epidemiological work for certain time-evolving variables
to be handled as covariates, as if the entire trajectory were recorded at baseline.
This approach is perfectly reasonable for an external variable such as air quality in an asthma study
where lack of cross-temporal interference might be defensible.
It has the advantage of leading to simple well-developed procedures for effect estimation
using marginal moments
(Zeger and Liang, 1986;
Zeger, Liang and Albert, 1988;
Laird, 1996;
Diggle, Heagerty, Liang and Zeger, 2002).
The same approach is less convincing for an evolving variable such as marital status in a survival study,
because the entire trajectory---suitably coded for $t>T_i$---would often contain 
enough information to determine the survival time.

\section{Survival prediction}
\subsection{Conditional distribution}
Consider the simplest model in which observations for distinct patients
are independent and identically distributed.
To simplify matters further, problems related to parameter estimation are set aside.
In other words, the survival time is distributed according to~$F$,
and the revival processes given $T=t$ is distributed as~$G(\cdot \given t)$.
Given the joint distribution, we are free to compute whatever conditional or marginal
distribution is needed to address the inferential target.

We consider here the question of how the partial trajectory of~$Y$
affects the subsequent survival prognosis.
The problem is to predict the survival time of an individual given
the survival process $Y[\bft^{(k)}]$ at the first $k$ appointments
 $\bft^{(k)}= (t_0<\cdots<t_{k-1})$. 

For positive real numbers $\bfs = (s_1> \cdots > s_k )$,
let $g_k(z; \bfs \given t)$ be the conditional joint density
given $T=t$ of the health-status values
\[
Z[\bfs] = (Z(s_1),\ldots, Z(s_k)) = (Y(T-s_1),\ldots, Y(T-s_k)).
\]
Under the conditional independence assumption (\ref{sci}),
which implies non-preferential appointment dates in the sense of
Diggle, Menezes and Su (2010), the joint density of 
$(T, \bft^{(k)}, Y[\bft^{(k)}])$ at $(t, \bft^{(k)}, y)$ is
a product of three factors:
\begin{eqnarray}
\nonumber
&& f(t) \times \prod_{j<k} p(t_j, y_j \given \H_j, T=t) \\
\nonumber
&=&f(t)\times \prod_{j<k} p(y_j \given \H_j, T=t) \times \prod_{j<k} p(t_j \given \H_j, T=t)\\
\label{jdensity}
&=&f(t) \times g_{k}(y; t-\bft^{(k)}\given t)
\times \prod_{j<k} p(t_j \given \H_j, T=t),
\end{eqnarray}
where $f=F'$ is the survival density, and $\H_j$ is the observed history
$(\bft^{(j)}, Y[\bft^{(j)}])$ at time $t_{j-1}$.
Without further assumptions, all three factors depend on~$t$,
meaning that all three components are informative for survival prediction.

In all subsequent discussion concerning prediction, it is assumed that
the appointment schedule is uninformative for prediction in the sense that
\begin{equation}\label{noninformative}
p(t_k \given \H_k, T=t) = p(t_k \given\H_k, T=\infty)
\end{equation}
for $t_{k-1} < t_k < t$.
This means that the next appointment is scheduled as if $T=\infty$,
but it is not recorded unless $t_k < T$.
With this assumption, the third factor in (\ref{jdensity}) is constant in~$t$
and can be ignored.
In other words, the distribution of the time to the next scheduled appointment
may depend on the patient's medical history, but
is independent of the patient's subsequent survival.
Ordinarily, the scheduled appointment is included as a component of the patient's
record only if it occurs in $[0, T)$ while the patient lives,
 implying that
the partial appointment schedule $\bft^{(k)}$ is uninformative for subsequent survival.
In particular, an administrative  schedule is uninformative.

\begin{figure}[ht]
\begin{center}
\includegraphics[height=6.0cm,width=12.0cm, angle=0]{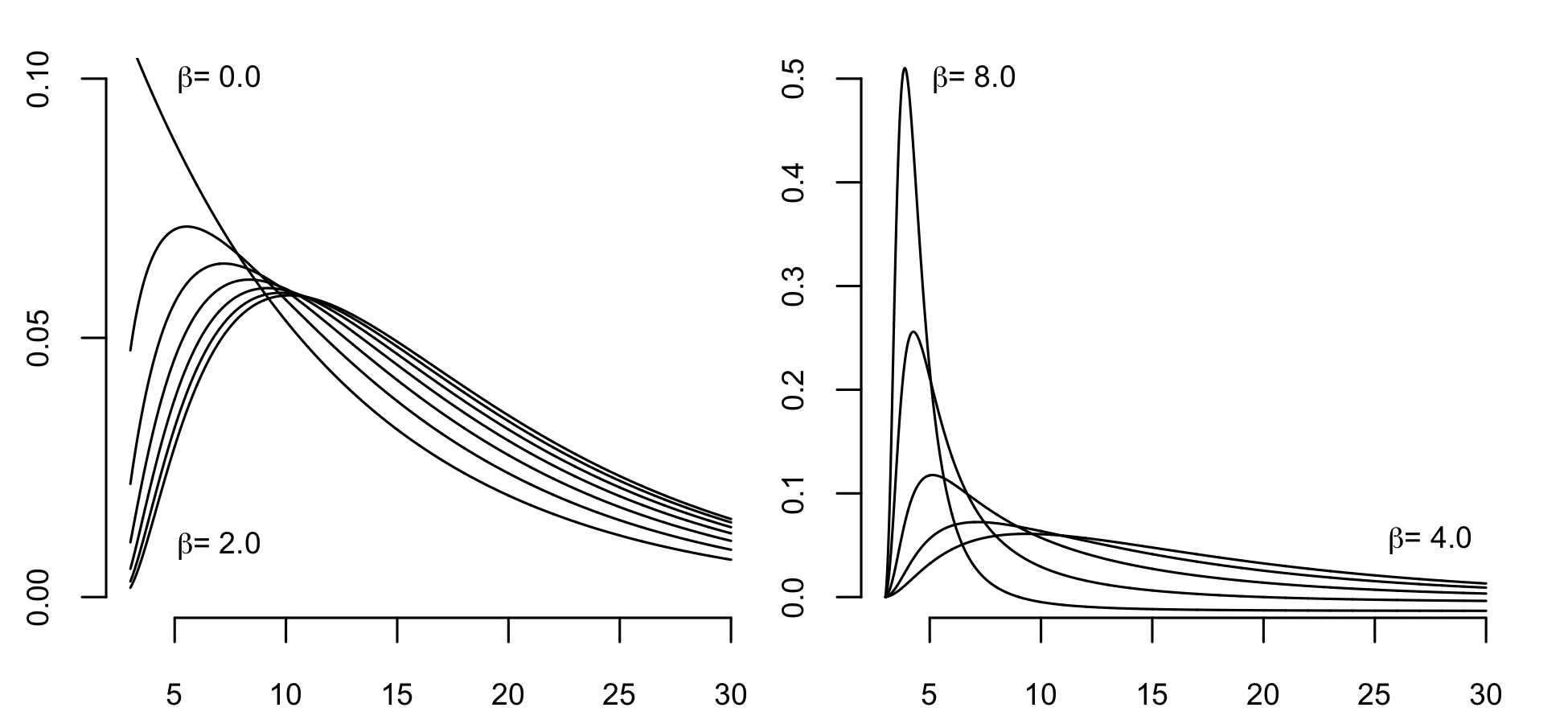}
\end{center}
\vspace{-0.5cm}
\caption{Conditional density of survival time for various values of $\beta$.}
\label{conditional_density}
\end{figure}

A simple numerical example illustrates the idea.
Suppose $T$ is exponentially distributed with mean 10 years,
and the revival process for $s > 0$ is a real-valued Gaussian process with
mean $E(Z(s)) = \beta s /(1+s)$ and covariance function
$\delta_{ss'} + \exp(-|s-s'|)$ for $s ,s'> 0$.
The observed health-status values at $\bft=(0, 1, 2, 3)$ are $y=(6.0, 4.5, 5.4, 4.0)$.

For $\beta=0$, the conditional density is such that
$T-3$ is exponential with mean~10;
the conditional density is shown for various values $0\le\beta\le 2$
in the left panel of Fig.~\ref{conditional_density},  and for $4\le\beta\le 8$ on the right.
Evidently, the conditional distribution depends on both the
observed outcomes and on the model parameters:
the median residual lifetime is not monotone in~$\beta$.
In applications where $\beta$ is estimated with appreciable uncertainty,
the predictive distribution is a weighted convex combination of
the densities illustrated.

The conditional survival distribution given $Y[\bft^{(k)}]$ depends not only on
the current or most recent value, but on the entire vector.
In particular, the conditional distribution does not have the structure of
a regression model in which the longitudinal variable enters as a time-dependent
covariate without temporal interference.
Thus, on the assumption that the joint model is adequate,
issues related to covariate confounding do not arise.

\subsection{A simple Gaussian revival process}
Under assumptions (\ref{sci}) and (\ref{noninformative}),
the ratio of the conditional survival density at~$t$ to
the marginal density is proportional to the factor
$g_k(y; t - \bft^{(k)})$, in which  $y, \bft^{(k)}$ are fixed, and $t$ the variable.
This modification factor---the Radon-Nikodym derivative---depends only on the
revival process, not on the distribution of survival times.
On a purely mathematical level,
it is precisely the likelihood function in the statistical model for the
$k$-dimensional variable $Y[\bft^{(k)}]$ whose conditional distribution
given $T=t$ is $G_k(y; t-\bft^{(k)} \given t)$
for some value of the temporal offset parameter~$t > t_{k-1}$.

%In a Gaussian revival model, both the mean vector
%$\mu[t-\bft]$ for fixed $\bft$,
%and the covariance matrix $\Sigma$ of $Y[\bft]=Z[t-\bft]$ may depend on~$t$.
%In the simpler circumstance where $\Sigma$ is independent of~$t$,
%$\log g$ is a quadratic function of $y - \mu[t-\bft]$,
%so the dependence of the likelihood on $t$ stems from the lack of constancy
%of the mean function.
%Clearly $g$ has its maximum at the value $\hat t\ge t_k$
%that minimizes $\normsq{y - \mu[t-\bft]}$ in the appropriate norm,
%which could occur at more than one internal point or at the extremes.
%For a locally stationary point, maximum or minimum,
%$\hat\mu' \Sigma^{-1}(y - \hat\mu) = 0$,
%where $\hat\mu = \mu[\hat t - \bft]$ and $\mu'$ is the derivative.

Although not realistic for most applications, suppose that $G$ is
Gaussian with mean $\mu(s) = \alpha + \beta s$
independent of $t$ and linear in reverse time,
and covariance function $\cov(Z(s), Z(s') \given t) = K(|s-s'|)$.
Then the log density ratio factor
\[
\log g_k(y; t-\bft^{(k)} \given t) = \hbox{const}
	-\vhalf (y - \mu[t - \bft])' K^{-1} (y - \mu[t - \bft]),
\]
is quadratic in~$t$ for $t > t_{k-1}$.
After substituting $\alpha + \beta(t - \bft)$ for the mean function,
and expressing the log density ratio as a quadratic in~$t$, it can be seen that
the predictive density ratio at $t > t_{k-1}$ is the density at $\beta t$ of the
Gaussian distribution with mean
\[
%-\one'\Sigma^{-1}(y - \alpha\one + \beta\bft)  \big/ ( \one'\Sigma^{-1} \one)
-\alpha + \one'K^{-1}(y + \beta\bft^{(k)}) \big/
	(\one'K^{-1}\one) = \bar y - \alpha + \beta\bar \bft
\]
and variance $1/(\one'K^{-1}\one)$, where $K$~has components $K(t_i - t_j)$.
Ignoring the dependence on the data that comes from parameter estimation,
the dependence of the predictive density ratio on the data for one patient comes 
through the weighted averages
\begin{equation}\label{weightedaverage}
\bar y = \one'K^{-1} y/(\one'K^{-1}\one),\qquad
\bar \bft = \one'K^{-1} \bft/(\one'K^{-1}\one)
\end{equation}
for this particular individual.

\subsection{Exchangeable Gaussian revival process}\label{exchangeable}
In a more natural Gaussian model, the revival processes for distinct
patients are exchangeable but not necessarily independent.
Revival models of this sort have much in common with growth-curve models
(Lee, 1988, 1991) in which
$Z_i(s) = \mu(s) + \eta_0(s) + \eta_i(s)$
is a sum of two independent zero-mean Gaussian processes,
and the mean function $\mu(s)$ is constant across individuals.
Usually the common deviation $\eta_0(\cdot)$ is moderately smooth but not stationary,
perhaps fractional Brownian motion with $\eta_0(0) =0$.
The idiosyncratic deviations are independent and identically distributed and they
 incorporate measurement error, so $\eta_i(\cdot)$ is ordinarily the sum of
a continuous process and white noise.
Thus, the Gaussian process is defined by
\begin{eqnarray}
\label{gaussianmean}
 E(Z_{is}) &=& \mu(s) \\
\nonumber
 \cov(Z_{is}, Z_{i's'}) &=& K_0(s,s') + \delta_{ii'}K_1(s,s')
	+  \sigma^2 \delta_{ii'}\delta_{ss'}
\end{eqnarray}
for some suitable covariance functions $K_0, K_1$, each of which
can be expected to have a variance or volatility parameter and a range parameter.
In the case of fractional Brownian motion, for example,
$K(s, t) \propto s^\nu + t^\nu - |s-t|^\nu$ for some $0 < \nu < 2$,
which governs the degree of smoothness of the random function.

For a new patient such that $Y[\bft^{(k)}] = y$, the conditional survival
density $\pr(T \in dt \given \hbox{data})$ given the data,
including the outcomes for the new patient, is computed in the same way as above.
The second factor in (\ref{jdensity})
is the density at the observed outcomes of the Gaussian joint distribution whose
means and covariances are specified above.
This involves all $n+1$ patients.

\subsection{Illustration by simulation}
\begin{figure}[ht]
\begin{center}
\includegraphics[height=6.0cm,width=12.0cm, angle=0]{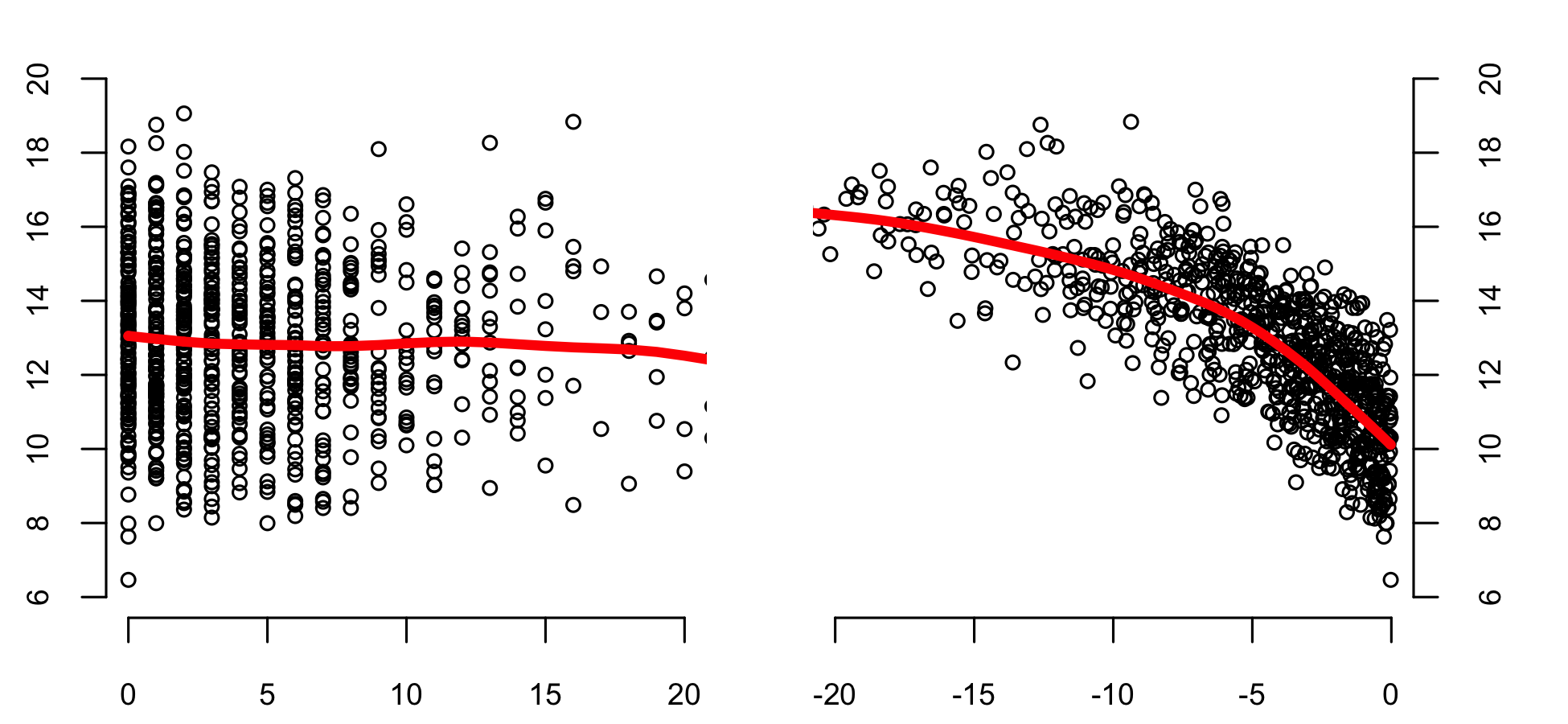}
\end{center}
\vspace{-0.5cm}
\caption{Simulated health status sequences aligned by recruitment  time (left) and 
	the same sequences aligned by failure time (right)}
\label{simulateddata}
\end{figure}

Figure~\ref{simulateddata} shows simulated data for 200 patients whose survival times are independent
exponential with mean five years.
While the patient lives, annual appointments are kept with probability $5/(5+t)$,
so appointment schedules in the simulation are not entirely regular.
Health status is a real-valued Gaussian process with mean $E(Z(s)) = 10 + 10 s/(10+s)$ in
reverse time, and covariances
$$
\cov(Z(s), Z(s')) = (1 + \exp(-|s-s'|/5) + \delta_{ss'}) / 2
$$
for $s,s' > 0$, so there is an additive patient-specific effect in addition to temporal correlation.
Values for distinct patients are independent and identically distributed.
This distribution is such that  health-status plots in reverse time aligned
by failure show a stronger temporal trend than plots drawn in the conventional way.
The state of health is determined more by time remaining before failure
than time since recruitment.
These trends could be accentuated by connecting successive dots for each individual,
as in Fig.~2 of Sweeting and Thompson (2011),
but this has not been done in Fig.~\ref{simulateddata}.

Since the survival times are exponential with mean five, independent of
covariates and treatment,  the root mean squared prediction error using covariates only is five years.
For fixed $k\ge 2$, and a patient having at least $k$ appointments, 
the conditional survival distribution given the first~$k$
health-status values has a standard deviation depending on the observed configuration,
but the average standard deviation is about $1.4$ years,
and the root mean squared prediction error is about $1.7$ years.
Using only the appointment schedule as a lower bound on the
survival time, the root mean squared prediction error is $3.9$ years.
For this setting, the longitudinal variable is a reasonably effective predictor of
survival, and the prediction error is almost independent of~$k$ in the range 2--5.  
This summary does not tell the full story because certain $y$-configurations
lead to very precise predictions whereas others lead to predictive distributions
whose standard deviation exceeds five years.

The parameter settings used in this simulation may not be entirely representative
of the range of behaviours of the conditional survival distribution given $Y[\bft]$.
If the ratio of the between-patient to within-patient variance components is increased, 
the average variance of the conditional survival distribution decreases noticeably with $k$.
For such settings, prediction using the entire health history is more effective
than prediction using the most recent value.

\section{Parameter estimation}\label{estimation}
\subsection{Likelihood factorization}
The joint density for the observations in a revival model
factors into two parts, one involving only survival times,
the other involving only the revival process.
%More generally, if the revival assumption fails, the second factor is the conditional
More generally, the second factor is the conditional
distribution of the revival process given~$T=t$, so both factors depend on~$t$.
Although both factors may involve the same covariates and treatment indicators,
the parameters in the two parts are assumed to be unrelated, i.e.,~variation independent.
Thus the likelihood also factors, the first factor involving
only survival parameters such as hazard modifiers associated with treatment and covariates,
the second factor involving only health-status parameters
such as temporal trends and temporal correlations.
In other words, the two factors can be considered separately,
either for maximum likelihood estimation or for Bayesian operations.

This approach is related to pattern-mixture modeling as discussed in 
Fieuws et al. (2008) in which the joint density $ \pr (T,Y)$ is
factorized as $\pr (Y \given T)  \pr (T)$.  Therefore the revival model 
can be viewed as a particular choice of pattern-mixture model.  
Initial contributions to the pattern-mixture approach include
Little (1993) in the context of longitudinal clinical trials with dropout.

\subsubsection{Survival distribution specification}

The first stage in parameter estimation is to
estimate the survival distribution $F$ together with
treatment and covariate effects if needed.
Whether the model for survival times is finite-dimensional or
infinite-dimensional, this step is particularly simple because
the first factor involves only the survival times and survival  distribution.
The standard assumption of independent survival times for distinct patients simplifies
the problem even further.
Exponential, gamma and Weibull models are all feasible,
with treatment effects included in the standard way.

For the Cox proportional-hazards model, the situation is a little more complicated.
First, the survival time is finite with probability one
if and only if the integrated hazard 
$\Lambda(\Real^+)= \int_0^\infty \lambda(t)\, dt$
is infinite, which is not satisfied at all parameter points in the model.
Second, the partial likelihood function depends only on
baseline hazard values $\lambda(t)$ in the range $0\le t \le \Tmax$,
where $\Tmax$~is the maximum observed survival time, censored or uncensored.
Thus, the likelihood does not have a unique maximum,
but every maximum has the property that
$\hat\lambda(t) = 0$ for all $0\le t \le \Tmax$ except for failure times,
at which $\hat\lambda$ has a discrete atom.
% in the homogeneous case, for any failure time $t$ strictly less than~$\Tmax$,
% $\hat\Lambda(\{t\})$ is the reciprocal of the risk set size at $t^-$
% (multiplied by the number of failures if tied failures are present).
% If $\Tmax$ is a failure time, $\hat\Lambda(\{\Tmax\}) = \infty$,
% implying that the estimated total hazard $\hat\Lambda(\Real^+)$ is infinite.
% Otherwise, if $\Tmax$~is a censoring time, $\hat\Lambda(\{\Tmax\}) = 0$
% and the estimated total hazard is finite.
By common convention (Kaplan and Meier 1958; Cox 1972, \S8)
$\hat\lambda(t) = 0$ for $t > \Tmax$,
but this choice is not dictated by the likelihood function.
Since the revival model requires survival times to be finite with probability one,
it is essential to restrict the space of
hazards to those having an infinite integral, which
rules out the standard convention for~$\hat\lambda$.
Equivariance under monotone temporal transformation points to
a mathematically natural choice
$\hat\lambda(t) = \infty$ for $t > \Tmax$;
a less pessimistic option is to use a finite non-zero constant such as
\begin{equation} \label{approx1}
\hat\lambda(t) = \frac{\hbox{total number of failures}}
		{\hbox{total person time at risk}}
\end{equation}
for $t > \Tmax$.
Both of these maximize the proportional-hazards likelihood
function---restricted or unrestricted---and
either one may be used in the revival model.

A less arbitrary alternative is to consider the set of {\it neutral to the right} processes (Kalbfleisch (1978), Clayton (1991), and Hjort (1990)).  Such processes are exchangeable survival process constructed by
generating survival times conditionally independent and identically distributed
via a completely independent hazard measure, i.e. the cumulative conditional
hazard is a L\'evy process.  These automatically satisfy the property that the survival
time is finite with probability one.  Dempsey \& McCullagh (2015) show a correspondence
with Markov survival processes, studying in particular the harmonic process for which 
the conditional distributions have a close affinity with the Kaplan-Meier estimator.  
For exchangeable survival times, the harmonic process is defined 
by two non-negative parameters, $(\rho, \nu )$.  The marginal survival time is
exponential with rate $\nu \cdot ( \psi ( 1 + \rho) - \psi ( \rho))$ where $\psi$ 
is the derivative of the log gamma function.  Given unique survival
times $T_1 < \ldots < T_k$ the conditional hazard is the product of a
continuous and discrete component.  The continuous component is
\[
H(t) = \sum_{i: T_i \le T} \nu \frac{T_{i} - T_{i-1}}{\Rsharp(T_{i-1}) + \rho} 
+ \nu\frac{T - T_{j}}{\Rsharp(T_j)+ \rho},
\]
where $\Rsharp (t)$ is the number of at risk individuals at time~$t-$. The sum runs over event times, censored or failure, such that $T_i \le T$,
and $T_j$ is the last such event.
The discrete component is a product over failure times
\begin{equation}\label{kaplan-meier}
\prod_{j : T_j \le t}  \frac{(\Delta^{d_j} \zeta)(r_j+1)}{(\Delta^{d_j} \zeta)(r_j)} =
\prod_{j : T_j \le t} \frac{r_j+\rho}{r_j+d_j+\rho}.
\end{equation}
%For small $\rho$, the discrete component is essentially the same as
%the right-continuous version of the Kaplan-Meier product limit estimator.    
%The complete conditional hazard function is then 
%\[
%\pr ( T_{n+1} > t | R[n] ) = \exp (- H(t) ) \prod_{j : T_j \le T} \frac{r_j+\rho}{r_j+d_j+\rho}
%\]
The hazard rate for $t > T_{\max}$ is constant, $\lambda = \nu / \rho$.  Given $\rho$, the maximum likelihood estimate for $\lambda$ is
\[
\left[ \frac{\rho}{k} \int_0^\infty ( \psi( \rho + \Rsharp (t) ) - \psi ( \rho) ) dt \right]^{-1}
\]
As $\rho$ tends to zero, the estimate tends to $k / T_{\max}$, while for $\rho \to \infty $, the estimate approaches equation~(\ref{approx1}).
Appendix~\ref{app_robustkm} derives the estimators as $\rho$ tends to zero when the marginal survival times are assumed
to be distributed Weibull.

The harmonic process has both a simple form for the joint density and is easy to generate sequentially.  Moreover, it is the only non-trivial Markov survival process with predictive distributions that are weakly continuous as a function of the initial configuration.  The only exception is the iid process, which arises as the limit $\rho \to \infty$ in which tied failures occur with probability zero.  Given the above, it is a natural choice when working with the revival process.

\subsubsection{Revival process specification}

The second stage, which is to estimate the parameters in the revival
process, is also straightforward, but only if all records
are complete with no censoring.
Serial dependence is inevitable in a temporal process, and there may also be independent
persistent idiosyncratic effects associated with each patient, either additive or multiplicative.
Gaussian revival models are particularly attractive for continuous health measurements
because such effects are easily accommodated with block factors for patients and
temporal covariance functions such as those included in the simulation in Fig.~\ref{simulateddata}.

Thus the second stage involves mainly the estimation of variance components
and range parameters in an additive Gaussian model.
One slight complication is that the revival process is not 
expected to be stationary, which is a relevant consideration in
the selection of covariance functions likely to be useful.
Another complication is that the health status may be vector-valued, $Y(t)\in\Real^q$,
so there are also covariance component matrices to be estimated.
If the covariance function is separable, i.e.
\[
\cov(Z_{ir}(s), Z_{ir'}(s')) = \Sigma_{r,r'} K(s, s')
\]
for some $q\times q$ matrix $\Sigma$,
maximum-likelihood estimation is straightforward.
But separability is a strong assumption implying that temporal correlations
for all health variables have the same pattern, including the same decay rate, 
which may not be an adequate approximation.
Nevertheless, this may be a reasonable starting point.

The second stage requires all health records to be aligned at their termini.
Accordingly, a record that is right censored ($T_i > c_i$) cannot be properly aligned.
If the complete records are sufficiently numerous,
the simplest option is to ignore censored records in the second stage,
on the grounds that 
the estimating equations based on complete records remain unbiased.
This conclusion follows from the fact that the second factor is the conditional
distribution given survival time. 
Thus, provided that the censoring mechanism is a selection based on patient survival time,
the estimating equations derived from complete records are unbiased.
The inclusion of censored records is thus more a matter of statistical efficiency than bias,
and the information gained from incomplete records may be disappointing 
in view of the additional effort required.

\subsection{Incomplete records} \label{incomp_imput}

%A record is said to be incomplete or right-censored if the survival time of the patient is
%unknown at the time of analysis. 
%It is inevitable in lengthy survival studies that some patients become lost to follow-up
%well before the trial is terminated. Censoring is more naturally regarded as an unavoidable artifact of the sampling scheme
%than as a part of the survival process.
%It must be accommodated in the analysis, but it is not in itself the primary focus.

If we choose to include in the likelihood the record for a patient censored at $c>0$,
we need the joint probability of the event $T > c$,
the density of the subset $\bft_c=\bft\cap[0,c]$,
and the outcome $Y[\bft_c]$ at $y$.
On the assumption that censoring is uninformative, i.e.,~that the distribution of the
subsequent survival time for a patient censored at time~$c$ is the same as
the conditional distribution given $T>c$ for an uncensored patient, the joint density is
\[
%\frac1{1-F(c)} \int_{t\ge c} f(t)\, p(\bft \given c, t) \, g(y; t - \bft)\, dt
\int_{t\ge c} f(t)\; p(\bft_c \given t) \; g(y; t - \bft_c)\, dt
\]
on the space of finite-length records.
Assumption (\ref{noninformative}) implies that the second factor,
the density of the appointment dates in $[0,c]$
for a patient surviving to time~$t>c$, does not depend on the subsequent
survival time~$t-c$, in which case it may be extracted from the integral.
It is also reasonable to assume that the distribution of appointment schedules is known,
for example if appointments are scheduled administratively at regular intervals,
in which case the second factor may also be discarded.
Since the survival probability $1-F(c)$ is included in the first-stage likelihood,
the additional likelihood factor needed in the analysis of the revival model is
\[
\frac1{1-F(c)} \int_{t>c} f(t)\, g(y; t-\bft_c)\, dt,
\]
in which $\bft_c$~may regarded as a fixed subset of $[0,c]$.
Unfortunately, the integral involves both the survival density $f(t) = F'(t)$ and the
density of the revival process, so the full likelihood no longer factors.
For an approximate solution, $f$~may be replaced with the estimate obtained
from the first-stage analysis of survival times, and if
$\hat f$ is purely atomic, the integral is converted to a finite sum.

For an incomplete record, the component of the derivative of the log-likelihood with respect to revival parameters, $\psi$, associated with the censored record $(Y[\bft^{(k)}], \penalty 0 \bft^{(k)}, c)$ is
\[
E_{\psi, \theta} \left[ \frac{ d \log g (y ; t - \bft_c; \psi )}{d \psi} \given (Y[\bft_c], \bft_c), T > c \right]
\]
where $\theta$ denotes the survival parameters assumed common to both. This is the expected value of the score given the observed censored record and censoring time.  Treating the survival time as missing data, a simple imputation method is proposed for approximate maximum likelihood estimation.  First, impute survival times, $T^\prime$, using the conditional survival distribution
\[
f \left( T \given ( Y[\bft_c], \bft_c), T > c ; \hat{\psi}_u, \hat{\theta}  \right)  \propto
f (T; \hat{\theta})\; g (Y[\bft_c]; T - \bft_c , \hat{\psi}_u ) \cdot {\bf 1} [ T > c]
\]
where $\hat{\psi}_u$ is the maximum likelihood estimate of the revival parameters for uncensored records, and $\hat{\theta}$ the maximum likelihood estimate of the survival parameters using both uncensored and censored records.

In this case the log-likehood component associated with the imputed, uncensored record is given by
\[
\log g (y ; T^\prime - \bft_c; \psi ) + \log f( T^\prime ; \theta),
\]
so parameter estimation after imputation is again straightforward.  Imputation performed multiple times creates imputed estimates $\{ \hat{\psi}^{(I)}_1, \ldots, \hat{\psi}^{(I)}_M \}$ with standard errors $\{ s^{(I)}_1, \ldots, s^{(I)}_M\}$.  These can then be averaged to get a complete-data estimate, $\hat{\psi} = \frac{1}{M} \sum_{m=1}^M \psi^{(I)}_m$.  A variance estimate, $V_{\psi}$ reflects variation within and between imputations:

\[
V_{\psi} = W + \left( 1+ \frac{1}{m} \right) B
\]
where $W = \frac{1}{M} \sum s_m^2$ and $B = \frac{1}{M-1} \sum_{m=1}^M \left( \hat{\psi}_m - \hat{\psi} \right)^2$.   Let $\hat{\psi}^{(c)}_{imp}$ denote the estimate for the censored records under imputation of the survival times.

Given maximum likelihood estimates, $\hat{\psi}^{(c)}$ and $\hat{\psi}^{(u)}$, and corresponding standard errors, $\hat{V}^{c}$ and $\hat{V}^{u}$, the following statistic is proposed for testing whether censored records are consistent with uncensored records:
\begin{equation} \label{teststatistic}
T_i = \frac{ \hat{\psi}_i^{(c)} - \hat{\psi}_i^{(u)} }{\sqrt{(\hat{V}_i^{c})^2 +(\hat{V}_i^{u})^2}}
\end{equation}
The denominator is the estimated variance of the difference under independence of the patients' revival processes.  While equation~(\ref{teststatistic}) can be used, when the survival times are imputed the estimates, $\hat{\psi}^{(u)}$ and $\hat{\psi}^{(c)}_{imp}$, are positively correlated as the imputed survival times use the maximum likelihood estimate for uncensored records, resulting in a conservative test statistic.  Appendix~\ref{tstat_appendix} discusses an appropriate modification of the test statistic in this case.

Exact likelihood analysis for incomplete records is technically more involved and is therefore omitted; however, the imputed estimates provide a first step in this direction.  The situation is considerably more complicated if, as in section~\ref{exchangeable},
the revival processes for distinct patients are not independent.

%For a revival parameter~$\theta$, a censored record contributes less information than
%an uncensored record of similar length.
%The log likelihood derivative generated by a complete record $(t, \bft, y)$ is
%\[
%U_\theta(t, \bft, y) = g'_\theta(y; t-\bft)/g_\theta^{}(y; t-\bft)
%\]
%where $g'_\theta$ is the derivative with respect to the parameter
%of the revival density.
%The log likelihood derivative for an incomplete record is the predictive expected value
%\[
%\bar U_\theta(c, \bft_c,y) = \int_{t>c} U_\theta(t,\bft_c,y) \, f_\theta(t \given Y[\bft_c]=y)\, dt.
%\]
%In other words, $\bar U_\theta = E(U_\theta \given Y[\bft_c]=y)$
%for fixed $\bft_c$ and $Y[\bft_c]=y$.
%The Fisher information measures are
%\begin{eqnarray*}
%\openup2\jot
%i(\theta) &=& \var(U_\theta(t, \bft, y)) \strut\cr
%\bar i(\theta) &=& \var \bar U_\theta(c, \bft_c, y)  = \var(E(U_\theta \given Y[\bft_c]=y))\strut\cr
%i(\theta) - \bar i(\theta) &=& E \var(U_\theta \given Y[\bft_c]=y),\strut
%\end{eqnarray*}
%where all moments refer to the joint distribution of $T, Y[\bft_c]$ for fixed $\bft_c$.
%It is not easy from these formulae to gauge the loss of information associated with censoring,
%partly because some parameters are affected more severely than others.

\subsection{Treatment effect: definition and estimation}
We consider here only the simplest sort of revival model for the effect of
treatment on patient health, ignoring entirely its effect on survival time.
Health status in the revival process is assumed to be Gaussian,
independent for distinct patients, and the treatment is assumed to have an
effect only on the mean of the process, not on its variance or covariance.
Consider two patients, one in each treatment arm,
\[
a_i(t) = \bar a_i(T_i - t) = 1, \qquad a_j(t) = \bar a_j(T_j - t) = 0
\]
such that $x_i = x_j$.
If $Z$ is independent of $T$, then the random variable $Z_i(s) - Z_j(s)$ is
distributed independently of the pair~$T_i, T_j$.
By definition, the treatment effect as defined by the revival model is the difference of means
\[
\tau_{10}(s) = E(Z_i(s)) - E(Z_j(s)) = E(Y_i(T_i-s)) - E(Y_j(T_j - s))
\]
at revival time~$s$.
This is not directly comparable with either of the the conventional definitions
\[
\gamma_{10}(t) = E(Y_i(t) - E(Y_j(t) ) \quad\hbox{or}\quad
\gamma'_{10}(t) = E(Y_i(t) - E(Y_j(t) \given T_i, T_j > t)
\]
in which the distributions are compared at a fixed time following recruitment.
The expectation in a survival study---that healthy individuals tend to live longer
than the frail---implies that $E(Y(t) \given T)$ must depend on the time
remaining to failure.
In that case, the conventional treatment definition $\gamma'_{10}(t)$
depends explicitly on the difference between the two survival times.
In other words, it does not disentangle the effect of treatment on
patient health from its effect on survival time.

%If $Z$ is not independent of $T$ then in the simplest 
%setting where the dependence on $T$ is linear and additive,
%the difference of means
%\[
%E(Z_i(s) \given T) - E(Z_j(s) \given T) = \tau_{10}(s) + \gamma(T_i - T_j).
%\]
%contains both a treatment effect and an effect due to the difference in survival times.
%In other words, the failure of the revival assumption does not necessarily
%complicate the interpretation of treatment effects.
If $Z$ is not independent of $T$ but the dependence  is additive,
 the difference of means at revival time~$s$
\[
E(Z_i(s) \given T) - E(Z_j(s)\given T) = \tau_{10}(s) + \gamma(T_i) - \gamma(T_j)
\]
contains both a treatment effect and an effect due to the difference in survival times.
In other words, the fact that $Z$ and $T$ are not independent does not necessarily
complicate the interpretation of treatment effects.
%
%The treatment effect in a revival process, may be persistent in time,
%i.e.,~constant for $s>0$ while the treatment is activated,
%but specifically excluding $s\ge \min(T_i, T_j)$, prior to randomization where the null level prevails.
%More complicated time-limited effects are also possible, but will not be considered here.
%More generally, there may be interactions in the sense that the
%magnitude of the treatment effect may depend on survival time,
%perhaps even negative for short-surviving patients and positive for long-surviving patients.
%
%For a single patient surviving to time~$T$, and an appointment schedule $\bft\subset[0,T)$
%equivalent to $\bfs=T - \bft$ in reverse time, the revival outcome $Y[\bft] \equiv Z[\bfs]$
%is distributed as
%\[
%Z[\bfs] \sim N(\mu[\bfs],\; K[\bfs])
%\]
%independently of~$T$.
%For simplicity,
%the covariance function $K$ is assumed to be independent of $x$ and treatment.
%Assuming that the treatment effect is constant across individuals,
%i.e.,~independent of~$x$, and also persistent over time, the revival mean has the form
%\[
%\mu_i(s) = \alpha(s, x_i) + \tau(\bar a_i(s))
%\]
%with a temporal trend $\alpha$ depending on~$x$, and an additive treatment effect~$\tau$.
%Various other forms of dependence, some simpler and others more complicated,
%can be specified using factorial model formulae in the standard manner---after
%the series have been aligned in revival order.
%
%Parameter estimates may be obtained by maximum likelihood using standard
%software for computing Gaussian likelihood functions.
%
By contrast with standard practice in the analysis of randomized trials with
longitudinal responses,
(Fitzmaurice, Laird and Ware 2011, section~5.6),
it is most unnatural in this setting to work with the conditional distribution
given the baseline outcomes $Y_i(0) \equiv Z_i(T_i)$.
This is one reason why the baseline response 
should be regarded as an integral part of the outcome sequence, not as a covariate.
Exchangeability implies distributional equality $Z_i(T_i) \sim Z_j(T_j)$ for individuals
having the same covariate values, but it does not imply equality of
conditional distributions given~$T$.
On the presumption that treatment assignment is independent of baseline response values,
we also have $Z_i(T_i)\sim Z_j(T_j)$ conditionally on treatment,
whether or not $a_i, a_j$ are equal.
Consequently, in order to satisfy the exchangeability assumption,
it is necessary to introduce a null, pre-randomization, treatment level,
$a_i(0) = a_j(0)$, common to all subjects.
%
%While treatment effects are not directly comparable, the revival effect can be translated to the original time scale.  In particular, assuming no observed history
%\[
%E ( Y_i (t) | T_i > t ) = \int_{0}^\infty E [ Z_i (s) ] \pr ( T_i = t+s | T_i > t) ds,
%\]
%So the mean of the conditional health process at time $t$ is a weighted average of the mean of the revival process.  Assuming a proportional hazards model, if the treatment effect on the survival distribution is constant, $\delta$ then $\gamma_{10}^\prime (t)$ can be expressed as 
%\[
%const + e^{\gamma} \int_{0}^\infty \tau_{10} (s) h_0 (t+s) \exp \left( - e^{\gamma} \int_{t}^{t+s} h_0 (x) dx \right)  ds
%\]
%where the constant depends on $\mu_0$.  In particular, if $h_0$ is constant then the conventional effect is a constant plus a weighted average of the revival effects.  Moreover, if $\delta = 0$ or $\mu_0 (t)$ is assumed to be $0$ for all $t$ then the conventional definition is exactly this weighted average.  

\subsection{Testing independence of $Z$ and $T$}
It is of interest to test whether the revival process~$Z$ is independent of
the survival time~$T$.  To do this, it is easy to formulate
and fit a specific alternative models in which the revival process is
\emph{not} independent of the survival time.
We consider here only the simplest design in which all records are complete,
there are no covariates or treatment assignment,
observations for distinct patients are independent,
and the revival model is a family of Gaussian process.
One way to do this is to replace (\ref{gaussianmean}) with
\[
E(Z_{i}(s) \given T) = \mu(s, T_i)
\]
for some suitable family of functions $\mu(s, T)$,
leaving the covariances unchanged.
For example, if $x$~denotes patient age at recruitment, the revival mean
might be modeled as
\[
E(Z_i(s) \given T) = \mu(s) + \beta_1 x_i + \beta_2 T_i
\]
depending additively on patient age and survival time.
If $\beta_1=\beta_2$, the dependence is on age at failure rather than
age at recruitment.
More general models involving multiplicative interactions between $s$ and $T_i$
may also be considered.

Consider, for instance, the non-linear Gaussian revival model with mean 
\[
\mu(s) = \alpha + \beta s/(\gamma + s),
\]
which is such that $\mu(0)=\alpha$,
$\mu(\infty)=\alpha+\beta$,
and $\mu(\gamma) = \vhalf(\mu(0) + \mu(\infty))$,
so that $\gamma > 0$ is the semi-revival time.
Within this family, the revival trajectory for one patient could be different from
that of another, depending on their survival times.
In other words, $\alpha,\beta, \gamma$ could depend on~$T$ or $x+T$, 
either of which is a violation of the independence assumption.
One of the simplest models of this type is the time-accelerated revival model
in which the semi-revival time is inversely related to survival,
\[
\mu(s, T) = \mu_0(sT) = \alpha + \beta s T/(\gamma + sT).
\]
As a practical matter, it would be more effective to replace $\gamma$ with
$\gamma_0 + \gamma_1/T$ or $\exp(\gamma_0 + \gamma_1/T)$ to
generate a test of independence.
Likewise, we could replace $\alpha$ with $\alpha_0 + \alpha_1 T_i$,
asserting that the outcome sequences for long-lived patients are elevated
by a constant amount at all revival times.
Similarly, if $\beta$ is replaced with $\beta_0 + \beta_1 T_i$,
the the asymptote is elevated in proportion to the additional lifetime.

Any modification of this sort is a violation,
so the survival time and the revival process are not independent.
However, the factorization of the likelihood function remains intact,
so the analysis remains relatively straightforward.
For example, a likelihood ratio statistic can %to test the revival assumption can
be constructed by fitting two nested models to the revival process, 
one assuming independence, the other not.%involving~$T$.

\section{A worked example: cirrhosis study}\label{cirrhosis_example}
\subsection{Prednizone and prothrombin levels}
In the period 1962--1969, 532 patients in Copenhagen hospitals
with histologically verified liver cirrhosis
were randomly assigned to two treatment arms,
control and prednisone. Only 488 patients for whom
the initial biopsy could be reevaluated using more restrictive criteria were 
retained, yielding 251 and 237 patients in the prednisone and placebo groups respectively.
Variables recorded at entry include sex, age, and several histological
classifications of the liver biopsy. Clinical variables were also collected, including
information on alcohol consumption, nutritional status, bleeding, and degree of ascites.
However, these covariates were not included in the dataset used here,
which was downloaded from the R~library {\tt http://cran.r-project.org/web/packages/joineR}
maintained by Philipson Sousa, Diggle, Williamson, Kolamunnage-Dona and Henderson.
%Pete Philipson, Ines Sousa, Peter Diggle, Paula Williamson, Ruwanthi Kolamunnage-Dona, Robin Henderson
At the end of the study period, the mortality rate was 292/488, or approximately 60\%.

\begin{figure}[ht]

\setbox0=\vbox{%
\includegraphics[height=5cm,width=6.5cm, angle=0]{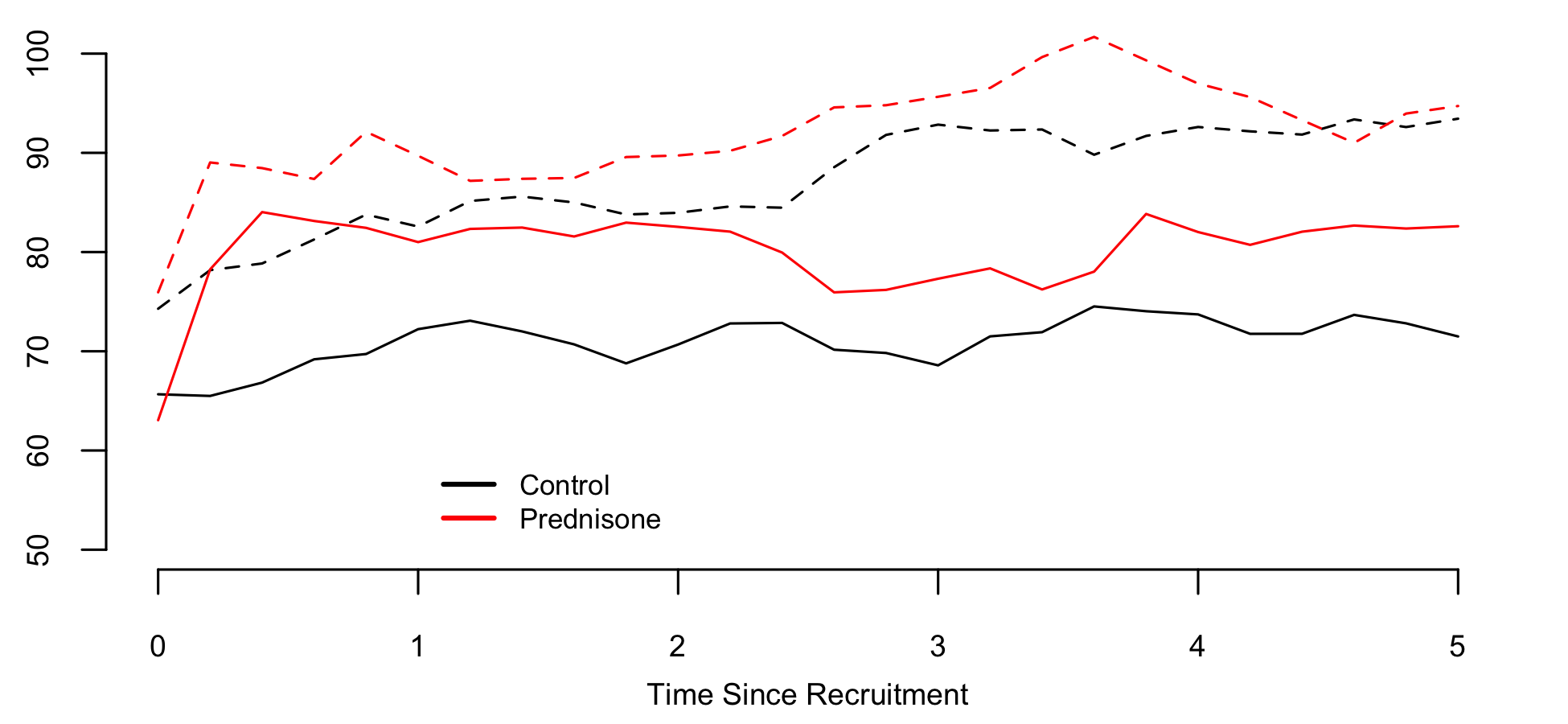}
}%
\setbox1=\vbox{%
\includegraphics[height=5cm,width=6.5cm, angle=0]{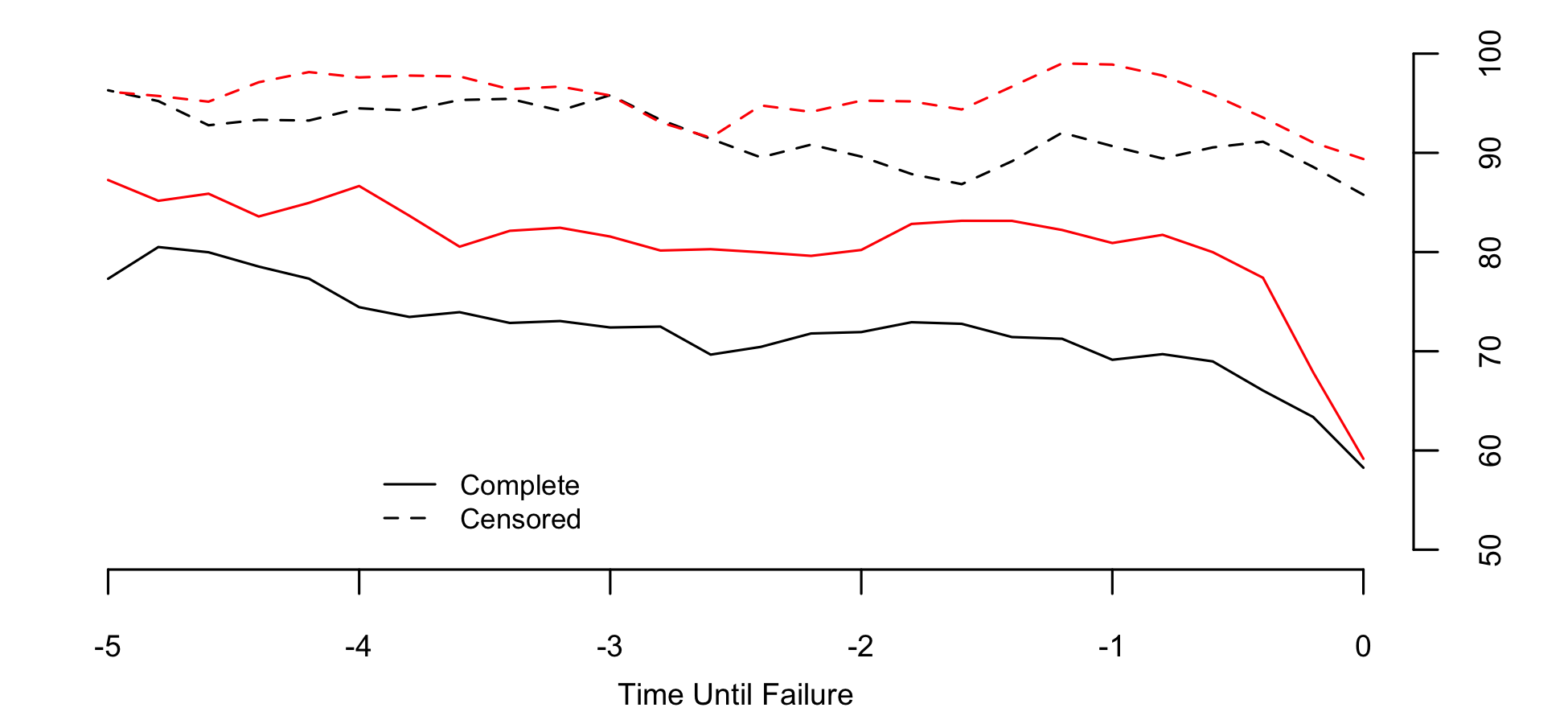}
}%
\begin{center}
\hbox to 19cm{\box0 \hss \box1}
\end{center}
\vspace{-1.0cm}
\caption{Prothrombin mean trajectories aligned by recruitment and by failure}
\label{fig:mean_traj}
\end{figure}
The focus here is on the prothrombin index, a composite blood coagulation index
related to liver function, measured initially at three-month intervals
and subsequently at roughly twelve-month intervals.
The individual prothrombin trajectories are highly variable,
both in forward and in reverse time,
which tends to obscure patterns and trends.
In Figure~\ref{fig:mean_traj}a the mean trajectory is plotted against time from recruitment
for two patient groups placebo/prednisone and censored/not censored.
Naturally, only those patients who are still alive are included in the average
for that time.
Figure~\ref{fig:mean_traj}b shows the same plots in reverse alignment.
While there are certain similarities in the two plots,
the differences in temporal trends are rather striking.
In particular, prothrombin levels in the six months prior to censoring are fairly stable,
which is in marked contrast with levels in the six months prior to failure,
as seen in the lower pair of curves.

Inspection of the graphs for uncensored patients in the right panel of Figure~\ref{fig:mean_traj} suggests
beginning with the simplest revival model in which the sequences 
for distinct patients are independent Gaussian with moments
\begin{eqnarray*}
E(Z_i(s) \given T) &=& \alpha + \tau_{\bar a_i(s)} + \beta_0 T_i + \beta_1 s +
	\beta_2\log(s+\delta) \\
\cov(Z_i(s), Z_j(s')\given T) &=& \sigma_1^2\delta_{ij}K_1(s, s') + \sigma_2^2\delta_{ij} +
	\sigma_3^2 \delta_{ij}\delta_{ss'}.
\end{eqnarray*}
The non-linear dependence on~$s$ is accommodated by the inclusion of $\log(s+\delta)$
in the mean model with a temporal offset~$\delta$, which is equal to one day
in all subsequent calculations.
Inclusion of the survival time $T_i$ is suggested by the increasing trend
along the diagonals and sub-diagonals of Table~1.
Since the value at recruitment is included as a response for each series,
treatment necessarily has three levels, {\sl null}, {\sl control\/} and {\sl prednisone}.
The three covariance terms are associated with independent additive processes,
the second for independent and identically distributed patient-specific constants,
and the third for independent and identically distributed white noise or measurement error.
The first covariance term governs the prothrombin sequences for individual patients,
which are assumed to be continuous in time with covariance function
$K_1(s, s') = \exp(-|s-s'|/\lambda)$ for $s, s' > 0$.
The temporal range in all subsequent calculations is set at $\hat\lambda=1.67$ years,
implying an autocorrelation of 0.55 at a lag of one year.
The implied one-year autocorrelation for the observed prothrombin sequences is
considerably smaller, roughly 0.30, because of the white-noise measurement term.

%> summary(fit1b)
%Likelihood kernel: K=(Intercept)+treatcontrol+treatprednisone+survival+revival+lrev
%Maximized log likelihood with kernel K is -5653.129 
%
%Mean-vector regression coefficients:
%                 Estimate Std. Error  Ratio
% (Intercept)      63.4557      2.023  31.37
% treatcontrol      2.4077      1.428   1.69
% treatprednisone  13.5580      1.474   9.20
% survival          1.7472      0.474   3.68
% revival          -2.1162      0.470  -4.50
% lrev              4.6624      0.407  11.45
%
%Variance coefficients: (after 3 cycles)
%                 Estimate Std. Error
%      Id         179.7105    12.9573
%      Patient    209.2155    34.9554
%      Patient.dt 211.4211    29.7746
%
For the initial likelihood calculations that follow, incomplete records are ignored;
only the 1634 measurements for the 292 non-censored patients are used.
The fitted variance components, estimated by maximizing the residual likelihood,
are 
\[
(\hat\sigma_1^2, \hat\sigma_2^2,\hat\sigma_3^2) = (210.0, 206.8, 179.6),
\]
all significantly positive.
Using these values to determine the covariance matrix,
the weighted least-squares coefficients in the mean model are shown in Table~\ref{coef_rev:tab}.
The standard error for the prednisone/control contrast is 1.77,
somewhat larger than the standard error for the prednisone/null contrast
because the former is a contrast between patients involving all three
variance components, whereas the latter is a contrast within patients,
which is unaffected by the second variance component.

\begin{table}[h]
\centering
\caption{Coefficients for revival model}
\label{tab:fit}
\begin{tabular}{l r r r c r r r r}
& \multicolumn{3}{c}{Censored Records} & & \multicolumn{3}{c}{Uncensored Records} \\ \cline{2-4} \cline{6-8} 
Covariate & Coef. & S.E. & Ratio &&  Coef. & S.E. & Ratio \\ \hline % & T\\ \hline
Null Treatment & 0.00 & - & - & & 0.00 & - & -  \\ % & -\\
Control & $4.13$ & 1.84 & 2.3 & & 2.41 & 1.43 & 1.7 \\ % & 0.98\\
Prednizone & $11.56$ & 1.75 & 6.6 & & 13.55 & 1.47 & 9.2 \\ % & $-1.20$\\
Survival ($T$) & 2.65 & 0.39 & 6.9 & & 1.75 & 0.47 & 3.7 \\ % & 2.67 \\
Revival ($s$) & $-2.78$ & 0.49 & $-5.7$ & & $-2.11$ & 0.47 & $-4.5$ \\ % & $-2.02$\\
$\log (s + \delta) $ & $3.74$ & 2.68 & 1.4 & & 4.66 & 0.41 & 11.5 \\ % & $-0.11$\\
\hline
$\lambda$ & & &  & & $0.164$ & \\ \hline % & \\ \hline
\end{tabular} 
 \label{coef_rev:tab}
\end{table}

Various deviations from this initial model may now be investigated.
In particular, it is possible to check whether there is an interaction
between treatment and survival time,
i.e.,~whether the treatment effect for long-term survivors is
or is not the same as the treatment effect for short-term survivors.
This comparison involves two variance-components models having
different mean-value subspaces, so the residual likelihoods are not comparable.
For likelihood comparisons, the kernel subspace must be fixed,
and the natural choice is the mean-value subspace for the null model
as described by Welham and Thompson (1997)
or as implemented by Clifford and McCullagh (2006).
The likelihood ratio statistic computed in this way is
0.83 on two degrees of freedom, showing no evidence of interaction.
However, there is appreciable evidence in the data that the
treatment effect (prednisone versus control) decreases 
as $t\to T$, i.e.,~as $s\to 0$.
The likelihood-ratio statistic for the $\hbox{\sl treat}.s$ interaction
is 3.90 on two degrees of freedom, showing little evidence of a linear trend,
but the value for the $\hbox{\sl treat}.\log(s)$ interaction
is 8.68, pointing to a non-linear trend.

We may also check the adequacy of the assumed form for the mean model
by including an additional random deviation, continuous in reverse time,
with generalized non-stationary covariance function such as
$K_0(s, s') = -|\log(s+\delta) - \log(s'+\delta)|$.
The fitted coefficient is 2.38, and the
associated likelihood ratio statistic is 1.2 on one degree of freedom,
showing no significant deviations that are continuous in reverse time.
Finally, we check whether the sequences for different patients exhibit
a characteristic pattern or trend associated with time measured from recruitment
by including the generalized Brownian-motion covariance function
$-|t-t'|$ in the covariance model.
The fitted variance coefficient is 2.10, and the likelihood ratio
statistic is 2.38 on one degree of freedom showing no significant characteristic patterns
that are continuous in time measured from recruitment.

Using the imputation method proposed in section~\ref{incomp_imput}, revival
parameters for censored records are estimated in order to check 
consistency with uncensored records.  Assuming the marginal
 survival time is exponential with rate parameter given by the first 
 stage maximum likelihood estimate, survival times are imputed. 
The imputation estimates are shown in Table~\ref{coef_rev:tab}.
Standard errors of coefficients that do not depend on the 
behavior near the origin of the revival times are similar to those for uncensored records.
Not surprisingly, the standard error for $\log (s+ \delta)$ is subtantially
higher for censored records.  The parameters associated with the survival and revival times show 
some deviation across record type, while treatment effects and the non-linear behavior 
with respect to the revival time appear consistent.  Conclusions appear robust to survival distribution specification as shown in Appendices~\ref{app_expfit}
and \ref{app_weibfit}, where imputed estimates under both exponential and Weibull
specifications for the above model as well as that including an interaction with 
treatment are provided.

A concern may be the parametric specification of the survival time distribution and 
whether this limits the method for handling censored records. 
To address this, Appendix~\ref{app_adjkmfit} shows estimates under the Markov survival process
specification when $\rho$ is sent to zero.  The result is a conditional survival distribution 
equivalent to the Kaplan-Meier product estimator for $t < T_{\max}$.  For $t \geq T_{\max}$
the hazard function is the Weibull hazard function.
Appendix~\ref{app_robustkm} finds maximum likelihood estimates for the survival distribution 
parameters when the marginal survival times are assumed Weibull. 
We see that the estimated conditional distribution is approximately 
equivalent to assuming the hazard is infinite for all
times after $T_{\max} = 13.40$.  Table~\ref{tab:kmfit} and~\ref{tab:kmintfit} 
shows the imputed estimates are similar to those under the exponential specification.

\subsection{Effect of prothrombin on prognosis}
Over a period of 5~years and one month following recruitment,
patient~$u$ had eight appointments with prothrombin values as follows:
\[
\arraycolsep=5pt
\begin{array}{rrrrrrrrr}
\hbox{$\bft_u$ (days)}  &0  &126  &226  &392  &770 &1127 &1631 &1855 \cr
\hbox{$Y_{u}[\bft_u]$}&49  &93 &122 &120 &110 &100  &72  &59 \cr
\end{array}
\]
This is in fact the record for patient $402$ who was assigned to prednisone
and was subsequently censored at 2661 days.
As determined on day~1855, the survival prognosis for this patient depends on
preceding sequence of measurements. 
Relative to the unconditional survival density for a patient on the prednisone arm,
the conditional survival density at time~$t>\max(\bft_u)$ is modified
multiplicatively by a factor
proportional to the joint conditional density of the random variable $Z_u[t-\bft_u]$
at the observed point~$y_u$ given $T_u=t$ and the data observed for all other patients.

\begin{figure}[ht]
\begin{center}
\includegraphics[height=6.0cm,width=12.0cm, angle=0]{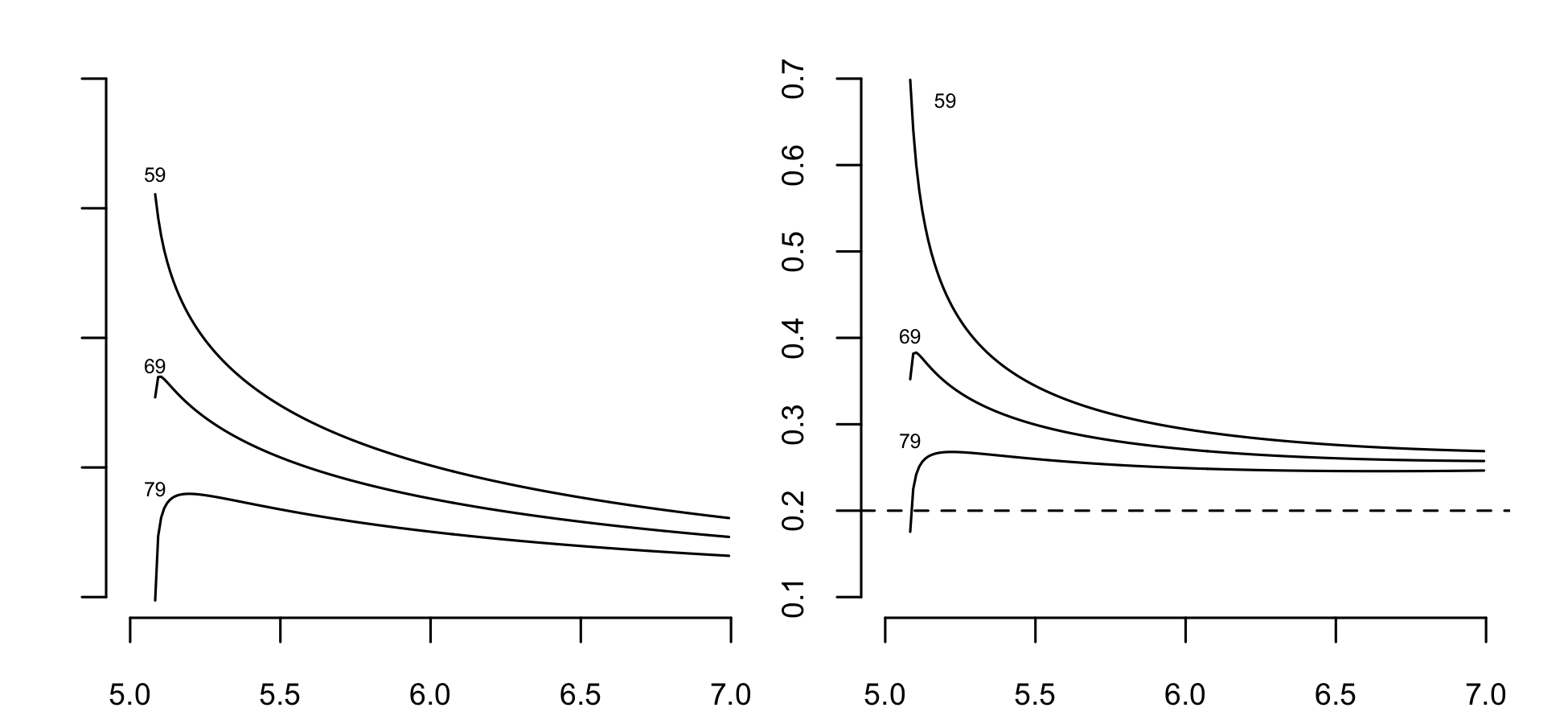}
\end{center}
\vspace{-0.75cm}
\caption{Three versions of the record for patient~402:
log modification factors for the predictive survival density (left panel) and
hazard functions (right panel).}
\label{predictivelogratio}
\end{figure}

For the model described in the preceding section---in which the records
for distinct patients are independent---this factor is particularly simple.
The conditional distribution of $Z_u[t-\bft_u]$ given $T=t$ has a mean vector
$\mu$ depending linearly on $t-\bft_u$ and $\log(t-\bft_u+\delta)$,
and a covariance matrix $\Sigma$ that is constant in~$t$.
The log density at $y_u$ is a quadratic form
\[
h(t, y_u) = \hbox{const} - (y_u - \mu)' \Sigma^{-1} (y_u - \mu)/2
\]
depending on~$t$ only through $\mu$.
This estimated factor is shown in Fig.~\ref{predictivelogratio}a for three versions of the record
in which the final prothrombin value is 59, 69 or 79.

It may be helpful to express the effect of the observed prednisone record
on the conditional survival distribution 
through its effect on the hazard function at times $t > \max(\bft_u)$
rather than its effect on the conditional survival density.
Suppose, therefore, that the unconditional survival time for
a patient on the prednisone arm, is exponential with mean 5 years,
so that the unconditional hazard function is constant.
What is the conditional hazard at time $t > \max(\bft_u)$ given
the prothrombin sequence for patient~$u$, with no further measurements
made in the interval $(\max(\bft_u), t)$ other than survival?
The conditional hazard functions for the subsequent two-year interval $5<t<7$
are shown in Fig.~\ref{predictivelogratio}b
for the same three versions of the prothrombin record.
It is evident from these plots that the conditional hazard for the real patient
is substantially elevated following the last measurement,
but the effect is transient and does not persist for the duration of a typical
inter-appointment interval of one year.
If the final value were 79 instead of 59, the hazard function is
almost constant, initially increasing and subsequently reverting
to the long-term value, which is slightly larger than the unconditional hazard.

The preceding analysis indicates that it may be misleading to treat
the observed health sequence as a time-dependent covariate in the proportional-hazards model.
At any one failure time $t$ measured from recruitment,
some of the health measurements are recent and fresh, while others
are likely to be up to one year old.
Figure~\ref{predictivelogratio}b shows that stale measurements may have negligible prognostic value
compared with fresh measurements.
The predictive revival model automatically takes into account the time that has
passed since the last appointment, so that stale values are discounted appropriately.

\subsection{Review of assumptions}
The conditional independence assumption (\ref{sci}) does not require appointments to be 
scheduled administratively, nor does it forbid patient-initiated appointments.
Consider two patients $i, j$ at time $s$ prior to failure, 
having similar prior appointment schedules and similar health values.
Assumption (\ref{sci}) states that the conditional appointment-initiation intensity
given the observed health record and subsequent survival time does not depend on 
subsequent health values.
In other words, conditional independence implies that patients $i, j$ are equally likely 
to initiate an appointment at time~$s$;
it is also assumed implicitly that they do so independently.

The evidence presented in section~\ref{schedules}, and in Liest\o l and Andersen (2002)
shows clearly that the rate of patient-initiated appointments increases in the last
few months of life.
It is certainly possible that patient behaviour in this instance violates the conditional independence assumption, but the evidence presented does not directly address the matter.
All in all, assumption (\ref{sci}) seems unavoidable and relatively benign.

The non-informative assumption (\ref{noninformative}) is much stronger than (\ref{sci}).
It implies that appointments are scheduled as if the patient will live indefinitely,
which is clearly contradicted by the evidence in this example.
We now examine the consequences of failure of (\ref{noninformative}), retaining (\ref{sci}).

Assumption (\ref{sci}) implies that the sampling is non-preferential in the sense of
Diggle, Menezes and Su (2010), which means that the second factor in (\ref{jdensity})
is the same as if the appointment dates had been fixed by design.
Consequently, the likelihood calculations in section~\ref{estimation} are unaffected 
by the failure of (\ref{noninformative}).

If the appointment for patient~$u$ on day 1855 were self-initiated in such a way that
the last factor in (\ref{jdensity}) depends on subsequent survival, 
it would be technically incorrect to omit that factor in prognosis calculations.
However, if it were known that all appointments for patient~$u$ were on schedule,
the possibility of a dependence on subsequent survival is eliminated,
and the prognosis calculations for this patient is technically correct
even if the behaviour of other patients violates (\ref{noninformative}).

\section{Summary}
The paper examines the problem of model formulation for health sequences,
whose defining characteristic is that the state space contains an absorbing value.
Each health sequence is terminated ultimately by death,
which is not equivalent to random restriction or censoring because subsequent values are known.
Typically, sequence length and sequence values are not independent.

The principal suggestion is that it may be more natural in some circumstances
to align health sequences by failure time than by age or by recruitment date.
The following list describes various statistical implications of realignment.
\begin{enumerate}
\parskip 1pt
\item{}
	The health sequence is regarded as a random process in
	its own right, not as a time-dependent covariate governing survival.
\item{}
	To a substantial extent, the model for survival time is decoupled
	from the revival model for the behaviour of the health sequence
	in reverse time.
\item{}
	Realignment implies that value $Y_i(0)$ at recruitment must not be treated
	as a covariate, but as an integral part of the response sequence.
	If they were available, values prior to recruitment could also be used.
\item{}
	The definition of a treatment effect is not the usual one
	because the natural way to compare the records for two individuals is not at
	a fixed time following recruitment, but at a fixed revival time.
	The treatment value need not be constant in revival time.
\item{}
	The predictive value of a partial health sequence for subsequent
	survival emerges naturally from the joint survival-revival distribution.
	In particular, the conditional hazard given the finite sequence
	of earlier values
	is typically not constant during the subsequent inter-appointment period.
\item{}
	Records cannot be aligned until the patient dies, which means
	that the revival process is not observable component-wise until $T$ is known.
	As a result, the likelihood analysis for incomplete records is
	technically more complicated.
	This aspect needs further development.
\item{}
	The omission of incomplete records from the revival likelihood does not
	lead to bias in estimation, but it does lead to inefficiency,
	which could be substantial if the majority of records are incomplete.
\item{}
	The principal assumption, that appointment dates be uninformative for subsequent survival,
	does not affect likelihood calculations,
	but it does affect prognosis calculations for individual patients.
	For that reason, it is advisable to label all appointments as scheduled or unscheduled.
\end{enumerate}

\acknowledgement
Comments by D.R.~Cox, D.~Farewell, R.~Gibbons, N.~Keiding, S.M. Stigler 
are gratefully acknowledged.

\def\journal#1{{\it #1}}
\def\title#1{{\it #1}}
\def\vol#1{{\bf#1}}
\def\pages{}
\section{References}
\everypar={\parskip 0.5pc \parindent 0pt \hangindent 10pt }
\noindent
Andersen, P.K., Hansen, L.H. and Keiding, N. (1991)
Assessing the influence of reversible disease indicators on survival.
\journal{Statistics in Medicine} \vol{10}, 1061-1067.

Clayton, D.G. (1991)
\title{A Monte Carlo method for Bayesian inference in frailty models.}
\journal{Biometrics} \vol{47}, \pages467--485.

Clifford, D. and McCullagh, P. (2006).
The regress function.
\journal{R Newsletter} \vol{6}, 6--10.

Cox, D.R. (1972)
Regression models and life tables (with discussion).
\journal{J.~Roy.\ Statist.\ Soc.}~B \vol{34}, 187--220.

Cox, D.R. and Snell, E.J. (1981)
\title{Applied Statistics}.
London: Chapman and Hall.

Dempsey, W. and McCullagh, P. (2014) 
Markov survival processes and proportional-hazards regression. 
Unpublished.

DeGruttola, V. and Tu, X.M. (1994)
Modeling progression of CD-4 lymphocyte count and its relation to survival time.
\journal{Biometrics} \vol{50}, 1003--1014.

Diggle, P.J., Heagerty, P., Liang, K.-Y. and Zeger, S.L. (2002)
\title{Analysis of Longitudinal Data}.
Oxford Science Publications: Clarendon Press.

Diggle, P.J., Farewell, D. and Henderson, R. (2007)
Analysis of longitudinal data with drop-out: objectives, assumptions and a proposal
(with discussion).
\journal{Applied Statistics} \vol{56}, 499--550.

Diggle, P.J., Sousa, I. and Chetwynd, A. (2008)
Joint modeling of repeated measurements and tome-to-event outcomes: The fourth Armitage lecture.
\journal{Statistics in Medicine} \vol{27}, 2981--2998.

Diggle, P.,  Menezes, R. and Su, T-L. (2010)
Geostatistical inference under preferential sampling (with discussion).
\journal{Appl. Statist.} \vol{59}, 191--232.

%Esbjerg, S., Keiding, N. and Koch-Henriksen, N. (1999)
%Reporting delay and corrected incidence of multiple sclerosis.
%\journal{Statistics in Medicine} \vol{18}, 1691--1706.

Farewell, D. and Henderson, R. (2010)
Longitudinal perspectives on event history analysis.
\journal{Lifetime Data Analysis} \vol{6}, 102--117.

Faucett, C.L. and Thomas, D.C. (1996)
Simultaneously modeling censored survival data and repeatedly measured covariates: a Gibbs sampling approach.
\journal{Statistics in Medicine} \vol{15}, 1663--1685.

Fieuws, S., Verbeke, G., Maes, B. and Vanrenterghem (2008)
Predicting renal graft failure using multivariate longitudinal profiles.
\journal{Biostatistics} \vol9, 419--431.

Fitzmaurice, G.M., Laird, N.M. and Ware, J.H. (2011)
\title{Applied Longitudinal Data Analysis, 2nd edition}.
New York: Wiley.

Fitzmaurice, G., Davidian, M., Verbeke, G. and Molenberghs, G. (2009)
\title{Longitudinal Data Analysis}
Chapman \& Hall.

Guo, X. and Carlin, B. (2004)
Separate and joint modeling of longitudinal and event time data using standard computer packages.
\journal{American Statistician} \vol{58}, 1--10.

Henderson, R., Diggle, P. and Dobson, A. (2000)
Joint modeling of longitudinal measurements and event time data.
\journal{Biostatistics} \vol1, 465--480.

Hjort, N.L. (1990)
\title{Nonparametric Bayes estimators based on beta processes in models for life history data.}
\journal{Annals of Statistics} \vol{18}, \pages1259--1294.

Kalbfleisch, J.D. (1978)
\title{Nonparametric Bayesian analysis of survival time data.}
\journal{J.~Roy.\ Statist.\ Soc.~B} \vol{40}, \pages214--221.

Kurland, B.F., Johnson, L.L., Egleston, B.L. and Diehr, P.H. (2009)
Longitudinal data with follow-up truncated by death:
match the analysis method to the research aims.
\journal{Statistical Science} 24, 211-222.

Lagakos, S.W. (1976)
A stochastic model for censored-survival data in the presence of an auxiliary variable.
\journal{Biometrics} \vol{32}, 551-559.  % a semi-markov model with intermediate states

Lee, J.C. (1988)
Prediction and estimation of growth curves with special covariance structures.
\journal{J.~Amer.\ Statist.\ Assoc.} \vol{83}, 432--440.

Lee, J.C. (1991)
Tests and model selection for the general growth curve model.
\journal{Biometrics} \vol{47}, 147--159.

Laird, N. (1996)
Longitudinal panel data: an overview of current methodology.
In {\it Time Series Models in Econometrics, Finance and Other Fields,}
D.R. Cox, D.V.~Hinkley and O.E.~Barndorff-Nielsen, eds.
Chapman \&~Hall Monographs on Statistics and Applied Probability 65.

%Lawless, J.F. and Nadeau, C. (1995)
%Some simple robust methods for the analysis of recurrent events.
%Technometrics 37, 158--168.
%
%Martinussen, T. and Scheike, T.H. (2006)	% relevant in principle
%Dynamic Regression Models for Survival Data.	% but only marginally
%Springer.

Liest\o l, K, and Andersen, P.K. (2002)
Updating of covariates and choice of time origin in survival analysis:
problems with vaguely defined disease states.
\journal{Statistics in Medicine} \vol{21}, 3701--3714.

Little, R.J.A. (1993).
Pattern-mixture models for multivariate incomplete data.
\journal{Journal of the American Statistical Association} \vol{88}, 125--134.

McCullagh, P. (2008).
Sampling bias and logistic models (with discussion).
\journal{J.~Roy.\ Statist.\  Soc.}~B \vol{70},  643--677.

Murphy, S.A. (2003)
Optimal dynamic treatment regimes (with discussion).
\journal{J.~Roy.\ Statist.\ Soc.}~B \vol{25}, 331--366.
%Molenberghs, G. and Verbeke, G. (2005)
%Models for Discrete Longitudinal Data.  Springer.
%
%Pepe, M. and Cai, J. (1993)
%Some graphical displays and marginal regression analysis for recurrent failure times
%and time-dependent covariates.
%J.~Amer.\ Statist.\ Assoc.\ 88, 811--820.

Rizopoulos, D. (2010)
JM: An R package for the joint modeling of longitudinal and time-to-event data.
\journal{Journal of Statistical Software} \vol{35}, 1--33.

Rizopoulos, D. (2012)
{\it Joint Models for Longitudinal and Time-to-Event Data}
Chapman and Hall.

Rosth\o j, S., Keiding, N. and Schmiegelow, N. (2012)
Estimation of dynamic treatment strategies for maintenance therapy of
children with acute lymphoblastic leukaemia: an application of
history-adjusted marginal structural models.
\journal{Statistics in Medicine}  \vol{31}, 470--488.

Schaubel, D.E. and Zhang, M. (2010)
Estimating treatment effects on the marginal recurrent
event mean in the presence of a terminating event.
\journal{Lifetime Data Analysis} \vol{16}, 451--477.

Sweeting, M.J. and Thompson, S.G. (2011)
Joint modeling of longitudinal and time-to-event data with application to predicting abdominal 
aortic aneurysm growth and rupture.
\journal{Biometrical Journal} \vol{53}, 750--763.

Tsiatis, A.A., DeGruttola, V. and Wulfsohn, M.S. (1995)
Modeling the relationship of survival to longitudinal data measured with error:
applications to survival and CD4 counts in patients with AIDS.
\journal{J.~Amer.\ Statist.\ Assoc.}\ \vol{90}, 27--37.

Tsiatis, A.A, and Davidian, M. (2004)
Joint modeling of longitudinal and time-to-event data: an overview.
\journal{Statistica Sinica} \vol{14}, 809--834.

van Houwelingen, H.C. and Putter, H. (2012)
\title{Dynamic Prediction in Clinical Survival Analysis}.
Monographs on Statistics and Applied Probability 123;
CRC Press.

Welham, S.J. and Thompson, R. (1997)
Likelihood ratio tests for fixed model terms using residual maximum likelihood.
\journal{J.~Roy.\ Statist.\ Soc.}~B \vol{59}, 701--714.

Wulfsohn, M.S. and Tsiatis, A.A. (1997)
A joint model for survival and longitudinal data measured with error.
\journal{Biometrics} \vol{53}, 330--339.

Xu, J. and Zeger, S.L. (2001)
Joint analysis of longitudinal data comprising repeated measures and times to events.
\journal{Applied Statistics} \vol{50}, 375--387.

Zeger, S.L. and Liang, K.-Y. (1986)
Longitudinal data analysis for discrete and continuous outcomes.
\journal{Biometrics} \vol{42}, 121--130.

Zeger, S.L., Liang, K.-Y. and Albert, P. (1988) 
Models for longitudinal data: a generalized estimating equation approach.
\journal{Biometrics} \vol{44}, 1049--1060.

\newpage 

\appendix

\section{Modification of test statistic} \label{tstat_appendix}

If the censored records are consistent with the uncensored records then given $\hat{\psi}^{(u)}$ the imputed parameter is approximately normal
\[ 
\hat{\psi}^{(I)}_j \; | \;  \hat{\psi}^{(u)} \sim^{a} N \left( \hat{\psi}^{(u)} , I \left( \hat{\psi}^{(u)} \right) \right)
\]
where $I \left( \hat{\psi}^{(u)} \right) = X_j^T \Sigma_j^{-1} X_j$ where the covariate matrix, $X_j$, and covariance matrix, $\Sigma_j$, are computed given the imputed survival time, $T_j$, at the parameter value $\hat{\psi}^{(u)}$. 

By the law of total variance, 
\begin{eqnarray}
\nonumber
\var_\theta \left( \hat{\psi}^{(c)}_{imp} \right) &=& E \left[ \var \left( \hat{\psi}^{(c)}_{imp} \;  | \; \hat{\psi}^{(u)} \right) \right] + \var \left( E \left[ \hat{\psi}^{(c)}_{imp} \;  | \; \hat{\psi}^{(u)} \right] \right) \\
\nonumber
&=& E_\theta \left[ X_{(c)}^T \Sigma_{(c)}^{-1} X_{(c)} \;  | \; \hat{\psi}^{(u)} \right] + X_{(u)}^T \Sigma_{(u)}^{-1} X_{(u)}
\end{eqnarray}
where $X_{(u)}$ and $X_{(c)}$ are the covariates for all uncensored and censored records respectively, and $\theta$ is the set of survival parameters. The law of total covariance implies the covariance is $X_{(u)}^T \Sigma_{(u)}^{-1} X_{(u)}$.  This implies that 
\[
\var \left( \hat{\psi}^{(u)}  - \hat{\psi}^{(c)}_{imp} \right)  = E_\theta \left[ X_{(c)}^T \Sigma_{(c)}^{-1} X_{(c)} \;  | \; \hat{\psi}^{(u)} \right] \approx W
\]
That is, the variance can be approximated from the standard errors of the imputed estimates.
Therefore for the imputed estimates, equation~(\ref{teststatistic}) is altered to
\[
T^\star_i = \frac{ \left(\hat{\psi}^{(c)}_{imp} - \hat{\psi}^{(u)}\right)_i}{\sqrt{W_{ii}}}
\]
for the~$i$th coordinate of the parameter vector.  Appendices~\ref{app_expfit} and \ref{app_weibfit} show the test statistic using this variant.

\newpage

\section{Estimates: marginal exponential survival} \label{app_expfit}

\begin{table}[!h]
\centering
\caption{Coefficients for revival model : no interaction}
\label{tab:fit}
\begin{tabular}{l r r r c r r r r}
& \multicolumn{3}{c}{Censored records} & & \multicolumn{3}{c}{Uncensored records} \\ \cline{2-4} \cline{6-8} 
Covariate & Coef. & S.E. & Ratio &&  Coef. & S.E. & Ratio & $T^\star$ \\ \hline
Null Treatment & 0.00 & - & - & & 0.00 & - & -  & -\\
Control & $4.13$ & 1.84 & 2.3 & & 2.41 & 1.43 & 1.7 & 0.94 \\
Prednizone & $11.56$ & 1.75 & 6.6 & & 13.55 & 1.47 & 9.2 & $-1.14$\\
Survival ($T$) & 2.65 & 0.39 & 6.9 & & 1.75 & 0.47 & 3.7 & 2.79 \\
Revival ($s$) & $-2.78$ & 0.49 & $-5.7$ & & $-2.11$ & 0.47 & $-4.5$ & $-1.72$\\
$\log (s + \delta) $ & $3.74$ & 2.68 & 1.4 & & 4.66 & 0.41 & 11.5 & $-0.46$\\
\hline
$\lambda$ & & &  & & $0.164$ & & \\ \hline
\end{tabular} 
\end{table}

\begin{table}[!h]
\centering
\caption{Variance components}
\label{tab:varcompfit}	
\begin{tabular}{l l r r c r r }
& & \multicolumn{2}{c}{Censored records} & & \multicolumn{2}{c}{Uncensored records} \\ \cline{3-4} \cline{6-7}
& & Coefficient & S.E. & &  Coefficient & S.E. \\ \hline
AR$1$ & $\sigma_1^2$ & 166.27 & 29.79 & &  209.95 & 29.54 \\
Patient & $\sigma_2^2$ & 155.84 & 31.02 & & 206.82 & 34.48 \\
White Noise & $\sigma_3^2$ & 223.69 & 17.30 & & 179.59 &  12.90 \\
\hline
\end{tabular} \\
\end{table}

\noindent\rule{5in}{0.4pt}

\begin{table}[!h]
\centering
\caption{Coefficients for revival model : with interaction}
\label{tab:fit}
\begin{tabular}{l r r r c r r r r}
& \multicolumn{3}{c}{Censored records} & & \multicolumn{3}{c}{Uncensored records} \\ \cline{2-4} \cline{6-8}
Covariate & Coef. & S.E. & Ratio &&  Coef.* & S.E. & Ratio & $T^\star$ \\ \hline
Null Treatment & 0.00 & - & - & & 0.00 & - & - & - \\
Control & $-1.49$ & 6.41 & $-0.23$ & & 1.79 & 1.51 & 1.19 & $-0.63$\\
Prednizone & $1.74$ & 5.07 & 0.34 & & 13.55 & 1.57 & 8.65 & $-2.50$\\
Survival ($T$) & 2.75 & 0.38 & 7.25 & & 1.78 & 0.48 & 3.70 & 2.98 \\
Revival ($s$) & $-2.73$ & 0.49 & $-5.57$ & & $-2.06$ & 0.49 & $-4.24$ & $-1.70$\\
$\log (s + \delta) $ & $0.17$ & 3.97 & 0.04 & & 4.07 & 1.06 & 3.85 & $-1.42$\\
$\log (s + \delta)$:Control & $2.46$ & 2.86 & $0.86$ & & $-0.31$ & 0.94 & $-0.33$ & 1.24 \\
$\log (s + \delta)$:Prednizone & $4.65$ & 2.25 & 2.07 & & 1.39 & 0.92 & 1.51 & 1.57 \\ \hline
\end{tabular} 
\end{table}

\begin{table}[!h]
\centering
\caption{Variance components for revival model : with interaction}
\label{tab:varcompfit}
\begin{tabular}{l l r r c r r }
& & \multicolumn{2}{c}{Censored records} & & \multicolumn{2}{c}{Uncensored records} \\ \cline{3-4} \cline{6-7} 
& & Coefficient & S.E. & &  Coefficient & S.E. \\ \hline
AR$1$ & $\sigma_1^2$ & 164.02 & 29.60 & &  212.31 & 29.53 \\
Patient & $\sigma_2^2$ & 155.57 & 31.54 & & 206.51 & 34.51  \\
White Noise & $\sigma_3^2$ & 223.23 & 17.27 & & 176.96 &  12.79 \\
\hline
\end{tabular} \\
\end{table}

\newpage
\section{Estimates : marginal Weibull survival} \label{app_weibfit}

\begin{table}[!h]
\centering
\caption{Coefficients for revival model : no interaction}
\label{tab:fit}
\begin{tabular}{l r r r c r r r r}
& \multicolumn{3}{c}{Censored records} & & \multicolumn{3}{c}{Uncensored records} \\ \cline{2-4} \cline{6-8} 
Covariate & Coef. & S.E. & Ratio &&  Coef. & S.E. & Ratio &  $T^\star$ \\ \hline
Null Treatment & 0.00 & - & - & & 0.00 & - & -  & -\\
Control & $4.11$ & 1.84 & 2.24 & & 2.41 & 1.43 & 1.69 & 0.93\\
Prednizone & $11.56$ & 1.75 & 6.61 & & 13.55 & 1.47 & 9.21 & $-1.14$\\
Survival ($T$) & 2.74 & 0.43 & 6.39 & & 1.75 & 0.47 & 3.70 & 2.89 \\
Revival ($s$) & $-2.77$ & 0.52 & $-5.73$ & & $-2.11$ & 0.47 & $-4.51$ & $-1.54$\\
$\log (s + \delta) $ & $3.29$ & 2.69 & 1.22 & & 4.66 & 0.41 & 11.47 & $-0.65$\\
\hline
$\lambda$ & & &  & & $0.159$ & & \\
$k$ & & &  & & $1.233$ & & \\ \hline
\end{tabular} 
\end{table}

\begin{table}[!h]
\centering
\caption{Variance components for revival model : no interaction}
\label{tab:varcompfit}
\begin{tabular}{l l r r c r r }
& & \multicolumn{2}{c}{Censored records} & & \multicolumn{2}{c}{Uncensored records} \\ \cline{3-4} \cline{6-7}
& & Coefficient & S.E. & &  Coefficient & S.E. \\ \hline
AR$1$ & $\sigma_1^2$ & 166.61 & 29.80 & &  209.95 & 29.54 \\
Patient & $\sigma_2^2$ & 156.17 & 31.14 & & 206.82 & 34.48 \\
White Noise & $\sigma_3^2$ & 223.70 & 17.30 & & 179.59 &  12.90 \\
\hline
\end{tabular} \\
\end{table}

\noindent\rule{5in}{0.4pt}

\begin{table}[!h]
\centering
\caption{Coefficients for revival model : with interaction}
\label{tab:fit}
\begin{tabular}{l r r r c r r r r}
& \multicolumn{3}{c}{Censored records} & & \multicolumn{3}{c}{Uncensored records} \\ \cline{2-4} \cline{6-8}
Covariate & Coef. & S.E. & Ratio &&  Coef.* & S.E. & Ratio & $T^\star$ \\ \hline
Null Treatment & 0.00 & - & - & & 0.00 & - & - & - \\
Control & $-1.97$ & 6.73 & $-0.29$ & & 1.79 & 1.51 & 1.19 & $-0.68$\\
Prednizone & $0.48$ & 5.57 & 0.09 & & 13.55 & 1.57 & 8.65 & $-2.57$\\
Survival ($T$) & 2.87 & 0.42 & 6.81 & & 1.78 & 0.48 & 3.70 & 3.14 \\
Revival ($s$) & $-2.68$ & 0.54 & $-4.94$ & & $-2.06$ & 0.49 & $-4.24$ & $-1.44$\\
$\log (s + \delta) $ & $-1.00$ & 4.47 & $-0.22$ & & 4.07 & 1.06 & 3.85 & $-1.63$\\
$\log (s + \delta)$:Control & $2.70$ & 3.11 & $0.87$ & & $-0.31$ & 0.94 & $-0.33$ & 1.23 \\
$\log (s + \delta)$:Prednizone & $5.41$ & 2.56 & 2.11 & & 1.39 & 0.92 & 1.51 & 1.73 \\ \hline
\end{tabular} 
\end{table}

\begin{table}[!h]
\centering
\caption{Variance components for revival model : with interaction}
\label{tab:varcompfit}
\begin{tabular}{l l r r c r r }
& & \multicolumn{2}{c}{Censored records} & & \multicolumn{2}{c}{Uncensored records} \\ \cline{3-4} \cline{6-7} 
& & Coefficient & S.E. & &  Coefficient & S.E. \\ \hline
AR$1$ & $\sigma_1^2$ & 164.18 & 29.58 & &  212.31 & 29.53 \\
Patient & $\sigma_2^2$ & 156.22 & 31.61 & & 206.51 & 34.51  \\
White Noise & $\sigma_3^2$ & 223.25 & 17.25 & & 176.96 &  12.79 \\
\hline
\end{tabular} \\
\end{table}

\newpage
\section{Estimates : adjusted Kaplan-Meier estimates} 
\label{app_adjkmfit}

\begin{table}[!h]
\centering
\caption{Coefficients for revival model : no interaction}
\label{tab:kmfit}
\begin{tabular}{l r r r c r r r r}
& \multicolumn{3}{c}{Censored Records} & & \multicolumn{3}{c}{Uncensored Records} \\ \cline{2-4} \cline{6-8} 
Covariate & Coef. & S.E. & Ratio &&  Coef. & S.E. & Ratio &  $T^\star$ \\ \hline
Null Treatment & 0.00 & - & - & & 0.00 & - & -  & -\\
Control & $4.06$ & 1.83 & 2.22 & & 2.41 & 1.43 & 1.69 & 0.90\\
Prednizone & $11.44$ & 1.75 & 6.53 & & 13.55 & 1.47 & 9.21 & $-1.21$\\
Survival ($T$) & 2.86 & 0.46 & 6.28 & & 1.75 & 0.47 & 3.70 & 2.71 \\
Revival ($s$) & $-2.82$ & 0.64 & $-4.42$ & & $-2.11$ & 0.47 & $-4.51$ & $-1.41$\\
$\log (s + \delta) $ & $3.58$ & 3.68 & 0.97 & & 4.66 & 0.41 & 11.47 & $-0.42$\\
\hline
\end{tabular} 
\end{table}

\begin{table}[!h]
\centering
\caption{Variance components for revival model : no interaction}
\label{tab:varcompfit}
\begin{tabular}{l r r c r r }
& \multicolumn{2}{c}{Censored Records} & & \multicolumn{2}{c}{Uncensored Records} \\ \cline{2-3} \cline{5-6}
& Coefficient & S.E. & &  Coefficient & S.E. \\ \hline
AR$1$ & 167.23 & 29.80 & & 209.96 & 29.54 \\
Patient & 158.08 & 30.57 & & 206.82 & 34.48 \\
White Noise & 223.08 & 17.28 & & 179.59 &  12.90 \\
\hline
\end{tabular} \\
\end{table}

\noindent\rule{5in}{0.4pt}

\begin{table}[!h]
\centering
\caption{Coefficients for revival model : with interaction}
\label{tab:kmintfit}
\begin{tabular}{l r r r c r r r r}
& \multicolumn{3}{c}{Censored Records} & & \multicolumn{3}{c}{Uncensored Records} \\ \cline{2-4} \cline{6-8}
Covariate & Coef. & S.E. & Ratio &&  Coef.* & S.E. & Ratio & $T^\star$ \\ \hline
Null Treatment & 0.00 & - & - & & 0.00 & - & - & - \\
Control & $-0.34$ & 8.42 & $-0.04$ & & 1.79 & 1.51 & 1.19 & $-0.38$\\
Prednizone & $3.57$ & 6.70 & 0.53 & & 13.55 & 1.57 & 8.65 & $-1.94$\\
Survival ($T$) & 3.10 & 0.52 & 5.98 & & 1.78 & 0.48 & 3.70 & 3.05 \\
Revival ($s$) & $-2.71$ & 0.70 & $-3.85$ & & $-2.06$ & 0.49 & $-4.24$ & $-1.28$\\
$\log (s + \delta) $ & $0.21$ & 5.62 & $0.04$ & & 4.07 & 1.06 & 3.85 & $-1.16$\\
$\log (s + \delta)$:Control & $1.98$ & 3.95 & $0.50$ & & $-0.31$ & 0.94 & $-0.33$ & 0.90 \\
$\log (s + \delta)$:Prednizone & $3.82$ & 3.10 & 1.23 & & 1.39 & 0.92 & 1.51 & 1.03 \\ \hline
\end{tabular} 
\end{table}

\begin{table}[!h]
\centering
\caption{Variance components for revival model : with interaction}
\label{tab:varcompfit}
\begin{tabular}{l r r c r r }
& \multicolumn{2}{c}{Censored Records} & & \multicolumn{2}{c}{Uncensored Records} \\ \cline{2-3} \cline{5-6} 
& Coefficient & S.E. & &  Coefficient & S.E. \\ \hline
AR$1$ & 165.21 & 29.69 & &  212.31 & 29.53 \\
Patient & 157.77 & 30.86 & & 206.51 & 34.51  \\
White Noise & 223.13 & 17.26 & & 176.96 &  12.79 \\
\hline
\end{tabular} \\
\end{table}

\newpage
\section{Robust estimation under adjusted nonparametric baseline hazard}
\label{app_robustkm}

As $\rho$ tends to zero, the discrete component of the 
conditional hazard for the harmonic process
converges to the Kaplan-Meier product limit estimator.
Unlike the Kaplan-Meier, for $\nu > 0$ the continuous component
is non-zero when $\Rsharp (t) > 0 $ and is undefined
for $t > T_{\max} = \max_{i} T_i$.    
On the other hand, if $\nu = \lambda \cdot \rho$ 
then the continuous component
is zero for $t < T_{\max}$, and equal to $\lambda$ for $t > T_{\max}$. 

Define ~$\nu(t) = \rho \cdot \lambda (t; \theta)$ so $\nu$ is time-dependent but 
proportional to $\rho$ at each time~$t$.  The marginal survival
time has distribution given by the hazard function~$\lambda (t) ( \psi(1+\rho) - \psi(\rho) )$.
The joint density is then given by
\[
 \frac{\prod_{i=1}^k \nu ( t_i; \theta) }{\rho^{\uparrow n}}
 \exp \left( 
   - \int_0^\infty \nu (s; \theta) \Psi ( \Rsharp (s) ) ds
 \right)
 \prod_{i=1}^k \Gamma ( d_i )
\] 
Assuming $\rho$ fixed, the log-likelihood as a function
of $\theta$ is
\[
  \sum_{i=1}^k \log ( \nu (t_i; \theta) )  
  - \int_0^\infty \nu (s; \theta) \Psi ( \Rsharp (s) ) ds
\]

As $\rho$ tends to zero,
the term $\rho \Psi ( \Rsharp(s) )$
tends to an indicator function of $s < T_{\max}$
so the log likelihood tends to
\[
\sum_{i=1}^k \log ( \lambda (t_i ; \theta) ) - \int_0^{T_{\max}} \lambda (s; \theta) ds
\]
We assume that the marginal distribution of each 
survival time is Weibull so that
\[
\lambda(t; \theta) = \frac{\kappa}{\lambda} \left( \frac{t}{\lambda} \right)^{\kappa - 1}
\]
In this case, the log-likelihood can be written as
\[
\sum_{i=1}^k \log ( \lambda (t_i; \theta) )  
  - \left( \frac{T_{\max}}{\lambda} \right)^\kappa = k \cdot [ \log(\kappa) - \kappa \, \log (\lambda) ] +  (\kappa - 1) \sum_{i=1}^k \log ( t_i ) - \left( \frac{T_{\max}}{\lambda} \right)^\kappa
\]

Differentiating with respect to $\lambda$ we have
\[
-\frac{k}{\lambda} + \frac{ T_{\max}^\kappa }{\lambda^{\kappa+1}} = 0
\]
which has solutions $\hat{\lambda} = 0$ and 
\[
\hat{\lambda} = \left[ k / T_{max}^\kappa \right]^{-1/\kappa}
\]
The first corresponds to the standard choice of
supposing a point mass at infinity, while the latter
to the non-zero maximum likelihood estimate
of interest.

Differentiating with respect to $\kappa$ we have
\[
  \frac{k}{\kappa} - k \log ( \lambda ) 
  + \sum_{i=1}^k \log (t_i) 
  - \left( \frac{T_{\max}}{\lambda} \right)^\kappa 
  \log \left( \frac{T_{\max}} {\lambda}\right)
  =
    \log (T_{\max})
    + \sum_{i=1}^k \log (t_i) 
    - \frac{k}{\kappa}
\]
solving for $\kappa$ we have 
\[
\hat{\kappa} = \left[ 
  \log (T) - \frac{1}{k} \sum_{i=1}^k \log (t_i)
\right]^{-1}
\]
The second term is the logarithm of the
geometric mean of the distinct survival
times.  Therefore, the estimator
is guaranteed to be greater than zero.
Moreover, the maximum likelihood estimate
is a simple function of the observed survival
and censoring times.

\subsection{Examples}

\subsubsection{Prednisone Case Study}

\begin{figure}
\label{fig:predsurv}
\centering
\includegraphics[scale = 0.5]{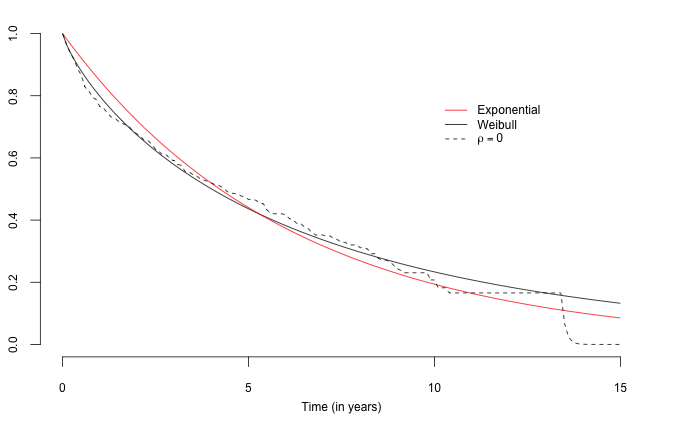}
\end{figure}

Applying the above estimators to the prednisone case study
we have
\[
\hat{\kappa} = 4.75 \times 10^{-1} \hspace{0.2cm} \text{and} \hspace{0.2cm} \hat{\lambda} = 1.05 \times 10^{-4}
\]
Figure~\ref{fig:predsurv} plots the survival curves using the maximum likelihood estimes
assuming the survival times are~i.i.d. exponential and Weibull distributed along with the 
kaplan meier estimator where the tail uses the above estimates.
The estimated survival curve is approximately equivalent
to assuming the hazard is infinite after the final observed time.

\subsubsection{Gehan Case Study}

\begin{figure}
\label{fig:gehan}
\centering
\includegraphics[scale = 0.5]{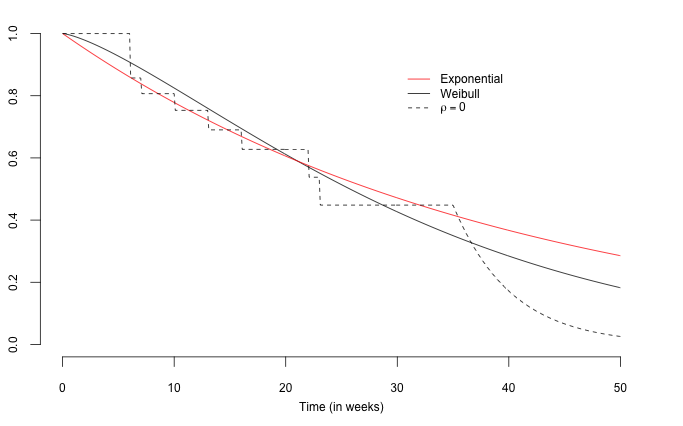}
\end{figure}

Consider parameter estimation for a set of failure and censoring times (in weeks) of
the 6-MP subset of leukemia patients taken from Gehan (1965):
\[
6,6,6,6^\star, 7, 9^\star, 10 , 10^\star, 11^\star, 13, 16 , 17^\star, 19^\star, 20^\star, 22, 23, 25^\star, 32^\star, 32^\star, 34^\star, 35^\star
\]
There are $9$ uncensored observations, and a total risk time of $359$ weeks.  
Applying the above estimators to the leukemia dataset
we have
\[
\hat{\kappa} = 0.96 \hspace{0.2cm} \text{and} \hspace{0.2cm} \hat{\lambda} = 4.62
\]
Figure~\ref{fig:gehan} plots the survival curves using the maximum likelihood estimes
assuming the survival times are~i.i.d. exponential and Weibull distributed along with the 
kaplan meier estimator where the tail uses the above estimates.
Here the estimated survival curve is not equivalent
to assuming the hazard is infinite after the final observed time.

\end{document}